\documentclass[alpha-refs]{wiley-article}

\usepackage{natbib}

\papertype{Research Article}

\title{Assessing the ability of a stretched-grid deep-learning weather prediction model to capture physical balances}

\newcommand{\authorfontsize}{\fontsize{11}{13.2}\selectfont}
\renewcommand{\Authfont}{\authorfontsize\bfseries}
\author[1,2\authfn{1}]{Francesco Pasquini}
\author[2]{Michiel Baatsen}
\author[1]{Bastien François}
\author[1]{Natalie Theeuwes}
\author[1]{Maurice Schmeits}

\affil[1]{Royal Netherlands Meteorological Institute (KNMI), De Bilt, The Netherlands}
\affil[2]{Institute for Marine and Atmospheric Research Utrecht (IMAU), Utrecht University, Utrecht, The Netherlands}

\corraddress{Francesco Pasquini, FBK, Via Sommarive 18, Trento, 38123, Italy.} 
\corremail{fpasquini@fbk.eu}

\presentadd[\authfn{1}]{Fondazione Bruno Kessler (FBK), Trento, Italy}

\runningauthor{Pasquini et al.}

\begin{document}

\maketitle

\begin{abstract}
Weather forecasting has traditionally relied on Numerical Weather Prediction (NWP) models, which simulate weather by solving the governing fluid equations. Recently, the emergence of Deep Learning Weather Prediction (DLWP) models has opened a new era in weather forecasting, offering a data-driven alternative to classical NWP approaches. Regional DLWP models such as the stretched-grid model Bris developed by Met Norway, have demonstrated performance on par with, or even slightly better than regional NWP models across a range of standard forecast metrics. By overcoming the coarse horizontal resolution that constrained earlier global data-driven models, the operational use of regional DLWP systems now appears increasingly promising. Nevertheless, the performance of such models during extreme events is generally inferior to that of regional NWP models, and comprehensive evaluations of their ability to generate physically realistic forecasts are still lacking. Here, we present a study comparing the physical consistency of the deterministic version of Bris with the control run of the operational MetCoOp Ensemble Prediction System (MEPS) in forecasting the severe extratropical cyclone Poly, which hit the Netherlands on 5 July 2023. We examine whether Bris accurately represents deviations from key atmospheric balances and whether it reproduces expected dynamics of the storm. We show that, despite its relatively good performance in terms of RMSE, Bris struggles to capture important mesoscale features of the event and that it significantly disrupts several atmospheric balances. This unrealistic disruption is mainly linked to the fine-scale noise evidenced in its output fields, which leads to incorrect and unrealistic spatial gradients. These results raise critical questions for improving AI-based models to better represent extreme events and how to ensure physical consistency in their predictions.

\end{abstract}

\vspace{0.2cm}

\section{Introduction}
In the last years, Machine Learning (ML) has become an extraordinarily powerful tool, with applications spanning from advanced scientific research to everyday life. When trained on large, high-quality datasets, ML-based models were seen able to recognize complex nonlinear patterns in data and make predictions in high-dimensional systems \citep[e.g.,][]{Battaglia2016}, with relatively low computational cost \citep{BenBouallegue2024}. Building on these capabilities, a rapidly growing line of research has emerged, aiming to develop fully data-driven models (DDMs) for weather forecasting. The goal is to progressively create a richer weather prediction environment by complementing traditional NWP approaches with new powerful instruments, helping to overcome some of their well-known limitations, such as the substantial computational time required for running numerical simulations.

A major breakthrough in this field came when \citet{Keisler2022} proposed a global weather forecasting model that employs a Graph Neural Network (GNN) trained on ERA5 reanalysis data to predict the 3D atmospheric state six hours ahead by learning spatio-temporal dynamics. The primary motivation for using a GNN was its demonstrated ability to model complex systems, with the expectation that the network could implicitly learn the physical laws represented in the ECMWF IFS model that underlies the ERA5 reanalysis dataset. Since then, a series of highly successful purely  DDMs has been developed, each introducing variations and improvements while remaining grounded on the same original idea proposed by Keisler. Notable examples include Pangu-Weather \citep{Bi2022}, GraphCast \citep{Lam2023}, and AIFS \citep{Lang2024}.

DDMs inherently benefit from the quality of numerical systems, as they are trained on their analyses, and have already demonstrated forecasting skill comparable to that of conventional physics-based NWP models for medium-range prediction. These results are especially remarkable considering that DDMs are still in their early stages, whereas NWP models have undergone decades of refinement and optimization. Moreover, DDMs can generate predictions at only a fraction of the real-time computational cost required by NWP-based systems \citep{Medina2025}. These considerable achievements led some researchers to regard it as the beginning of a new era in weather forecasting and motivated the development of higher-resolution DDMs for domains within Europe. A first example is Bris, developed by MET Norway \citep{Nipen2024}, a stretched-grid DDM that provides enhanced resolution over a selected region while maintaining a coarser grid elsewhere. Its performance achieves lower RMSE for several surface variables compared with the state-of-art Harmonie-based NWP model used in Norway, which shares the same horizontal resolution.

However, despite the remarkable performance of Bris and other global DDMs in terms of standard skill metrics (e.g., RMSE), open questions remain regarding their ability to generate realistic and physically consistent predictions. In particular, it is still unclear whether these models produce forecasts that are not only statistically skilful but also capture the underlying physical relationships present in the training data, as proposed in \citet{Keisler2022}. Some researchers have explicitly examined this issue \citep{Bonavita2024, Hakim2024}, but the answer has not yet been thoroughly investigated for a stretched-grid DDM such as Bris.

Studies conducted at coarser resolutions have reached contrasting conclusions. Some suggest that DDMs fail to adequately reproduce sub-synoptic and mesoscale phenomena and lack the fidelity and consistency  of physics-based models \citep{Bonavita2024}, whereas others indicate that these models may indeed encode atmospheric dynamics \citep{Hakim2024}. Answering this question has become increasingly urgent, as national meteorological services consider it a fundamental prerequisite before integrating such systems into operational forecasting workflows. Physical consistency is in fact widely regarded as a proxy for a model’s realism, a characteristic that in operational meteorology it may even take precedence over purely statistical accuracy, especially in extreme weather.

This represents a significant obstacle to the adoption of deterministic DDMs in real-world forecasting. These models often struggle to accurately reproduce extreme events, frequently underestimating their magnitude against observations \citep[e.g.,][]{Wijnands2025}. Among the factors that can explain this underperformance are the use of MSE as a loss function, which favours spatial smoothing to mitigate the double-penalty problem \citep{Subich2025}, the limited exposure during training to such rare extremes \citep{Olivetti2024}, and the inability to capture the underlying dynamical processes driving such events \citep{Pasche2024}.

This study evaluates the deterministic version of Bris (MSE-Bris \citep{Nordhagen2025}, which we will refer to simply as Bris for brevity), with particular emphasis on these two key challenges faced by DDMs: the ability to respect physical laws and the skill to forecast extreme events. A first central question we address is whether Bris, which is trained on time series of analyses grounded in a physically simulated background, can generate outputs that preserve expected atmospheric balances. To this end, we examine whether departures from these balances are consistent with those found in NWP forecasts and in the corresponding analysis fields. This approach builds on the methodology of \citet{Bonavita2024}, extending it to a higher-resolution regional model (2.5 km) and to a broader set of atmospheric balances.

The second critical question is determining whether such a high-resolution DDM can capture high-impact features of an intense windstorm, an essential aspect for assessing the model’s readiness for operational use. By focusing on an episode particularly severe like Poly, of which the occurrence is either rare or non-existent in Bris’s training dataset, our analysis targets the model’s intrinsic ability to generate realistic atmospheric patterns rather than its tendency to reproduce familiar situations encountered during training. By analysing the physical consistency of Bris and its capabilities under extreme weather we hope to clarify its potential and its current limitations, and pinpoint possible improvements to make such regional DLWP models ready to be used in operations.

\section{Models and Case Study}

The forecasts of Bris \citep{Nipen2024} are compared with those from the control run of the MetCoOp Ensemble Prediction System (MEPS) \citep{Andrae2020}, which is the NWP system employed by MET Norway and other countries in Northern Europe, and with the corresponding analysis fields. The comparison with MEPS is motivated by the fact that MEPS runs at the same horizontal resolution of 2.5 km \citep{Andrae2020} and that MEPS analyses have been used for the training of Bris. We focus exclusively on the MEPS control run rather than the ensemble mean, because at the time this study was conducted, only a deterministic version of Bris was available (probabilistic versions of the model, Bris CRPS and Bris CRPS-FFT have only recently been introduced \citep{Nordhagen2025}). This choice also aligns with our goal of assessing the physical consistency of individual model outputs, since ensemble means, while statistically useful, are not physically realistic states of the atmosphere.

Bris learns its prediction function directly from data. It is first pre-trained on 43 years of ERA5 reanalysis (1 January 1979 – 31 May 2022) at 31 km grid spacing, with aim to enable the model to capture robust large-scale atmospheric dynamics and synoptic variability \citep{Nipen2024}. The model is subsequently fine-tuned on a 3.3-year combined dataset (6 February 2020 – 31 May 2023), consisting of IFS analyses interpolated to the ERA5 resolution and MEPS analyses over the regional domain at 2.5 km grid spacing. Model performance is then evaluated over an independent test period from 1 June 2023 to 31 May 2024 (see \citep{Nipen2024} for further details). 

In contrast, the MEPS control run is based on the Harmonie–Arome model \citep{Andrae2020}, which predicts future atmospheric states by integrating the non-hydrostatic, terrain-following, fully compressible Euler equations forward in time \citep{Gleeson2024}. In this NWP model, simulations are initialized using the corresponding MEPS analyses.

A central point of this evaluation is that MEPS forecasts are based on physical equations, and we therefore do not expect the Navier–Stokes equations or the associated atmospheric dynamical balances to break down substantially. Similarly, MEPS analyses are obtained by combining a numerical short-term forecast with the assimilation of real observations and large-scale mixing processes \citep{Muller2017}, and therefore these fields should also remain close to the expected atmospheric balances. Bris, on the other hand, is not constrained by physical laws. As a result, it could violate key atmospheric balances and produce spatial patterns that deviate from those found in the NWP forecasts and analysis fields, despite being trained on physically-based analysis data.

MEPS and Bris share the same high-resolution regional domain, encompassing the entire Scandinavian region and surrounding areas, including nearly the entire Netherlands \citep{Andrae2020}. Windstorm Poly, driven by the strong westerlies typical in these mid-latitudes, entered the southwestern corner of this regional domain around 00UTC on 5 July 2023. Its rapid cyclogenesis was initiated by the interaction between a deepening upper-level trough and a localized convective layer above Brittany \citep{EUMETSAT}. After emerging over the North Sea, the cyclone’s north-eastward motion slowed considerably, allowing it to intensify over the unusually warm waters. Poly reached full development at around 06UTC as it approached the Dutch coast and, by 18 UTC, its core moved towards Denmark, exhibiting a reduced intensity.

In addition to falling within the Bris' testing period, Poly was selected as a case study due to its particularly challenging predictability and its significant impact on the Netherlands, highlighting its operational relevance for the Dutch weather service (KNMI). Previous studies have shown that capturing Poly’s dynamics requires high-resolution data, as coarser reanalyses such as ERA5 fail to show key processes \citep{Kees2024}. This makes Poly an ideal test case for evaluating Bris’s ability to generate realistic high-resolution forecasts. Moreover, as an extratropical cyclone, Poly provides a useful framework for studying fundamental atmospheric dynamics using a limited set of variables, a practical advantage given the restricted set of meteorological variables available from Bris (e.g., for examining gradient-wind balance using only geopotential and wind).

For a more detailed description of Poly’s characteristics and dynamics, the reader is referred to \citet{Kees2024}. Extensive information on the deterministic Bris model configuration can be found in \citet{Nipen2024}, while further details on the MEPS model and its analyses are provided in \citet{Muller2017} and \citet{Andrae2020}.

\section{Data and Methodology}
\label{sec3}

This study is based on three distinct datasets: (i) the DDM predictions produced by Bris, (ii) the NWP forecasts generated by the control member of MEPS, and (iii) the corresponding analysis fields. MEPS forecasts and their associated analyses are publicly available through the MEPS archive \citep{MEPSarchive}. The Bris output was provided by MET Norway and covers the period 1–12 July 2023, a 12-day window encompassing the full life cycle of the extratropical cyclone Poly, from its initial development to several days after its dissipation.

The analyses, used both as initial conditions for MEPS and Bris, and as training labels for Bris, serve as our reference fields, as they provide the spatio-temporal patterns Bris is expected to learn during training. However, when evaluating departures from physical balances, our aim is not to determine which model aligns more closely with this reference. Analyses are not necessarily expected to be strictly physically consistent, in contrast to NWP forecasts, as the goal of data assimilation is to bring the atmospheric state closer to observations, which can lead to small deviations from the governing dynamical equations. Instead, our objective is to assess whether Bris exhibits substantially larger deviations from atmospheric balances compared with both MEPS forecasts and the analyses—two physically constrained systems that, in principle, should behave similarly.

The evaluation is conducted over the high-resolution regional domain common to Bris and MEPS, excluding both the coarser global model domain and the outermost eight grid points ($\approx$ 20 km) of the regional domain that form the transition zone between the regional and global domains. This exclusion is motivated by the observed presence of imbalance outliers near the lateral boundaries of the MEPS and analysis regional domains, which arise from resolution-transition effects inherent to nested NWP systems. An analysis of balance disruptions within this transition zone, while potentially of interest, lies beyond the scope of the present study and such effects are left for future investigation.

As Bris is evaluated during the passage of windstorm Poly, we restrict most of our analysis to the \textit{south-western portion} of this Bris–MEPS regional domain, an area over which the storm’s Low-Pressure Centre (LPC) tracked during its mature stage and the impact of the event was most pronounced. Specifically, after the exclusion of the outermost eight grid points, the selected area under examination is a square domain of 443 $\times$ 443 grid points ($\approx$1100 km × 1100 km), centred roughly over the eastern part of the North Sea (see Fig. S1). Finally, for the assessment of atmospheric balances, we focus to the upper atmosphere. 

These methodological choices allow the study not only to extend previous work, where the physical consistency was typically examined for global DDMs with coarser resolutions \citep[e.g.,][]{Bonavita2024,Hakim2024}, but also to explore a region of the atmosphere (the \textit{upper atmosphere}) where Bris has so far received limited attention. While earlier research addressed its performance on surface variables \citep[e.g.,][]{Nipen2024}, those which are most heavily weighted in the loss function, our investigation gives also importance to high-altitude variables that carry less weight in the training process. Furthermore, analyzing dynamical balances in the middle and upper troposphere is facilitated by the dominance of large-scale weather patterns, the tendency of the atmosphere to remain near quasi-equilibrium, and the smaller influence of topography features such as mountains.

The limited temporal resolution (6-hourly) and the restricted set of output variables available from the DDM were main obstacles for this analysis (input/output variables of Bris are shown in Table 1 of \citet{Nipen2024}). These data constraints strongly shaped the design of the study, requiring us to make fundamental assumptions and to simplify the physical representation of the storm. For example, the 6-hour time step prevented any meaningful computation of time derivatives, making it impossible to investigate the temporal evolution of the meteorological variables in detail. Consequently, the study focuses on relationships that assume dynamical equilibrium, where local acceleration is negligible. Although this is a strong approximation, its implications are carefully considered when interpreting the results. 

We also outline a general procedure adopted throughout the analysis of multiple physical relationships. To assess the physical consistency of the model at different spatial resolutions, we first compute and evaluate the imbalances using the raw (unsmoothed) fields. The same diagnostics are then repeated after applying an artificial spatial smoothing. This step plays a crucial role in the present investigation, as will become clear in Section~\ref{geo_bal_section}. The degree of smoothing is indicated by \textit{factor} \( N \) (\( fN \)), where \( N \) denotes that the value at each grid point is replaced by the average over an \( N \times N \) grid-point window. For example, \( f1 \) corresponds to no smoothing, whereas \( f7 \) represents smoothing over a \( 7 \times 7 \) grid-point area (approximately \( 17.5 \times 17.5\,\mathrm{km}^2 \)). 

In the following sections, we describe step by step the methodology adopted to examine each dynamical balance. Specifically, we consider the hydrostatic, geostrophic, and gradient wind balances, together with an assessment of the continuity equation. This set of balances does not constitute a complete system for a standard verification of the physical consistency of a weather prediction model. Rather, it reflects the subset of dynamical relationships that can be meaningfully investigated given the limited availability of output variables from the Bris model and, most importantly, its coarse temporal resolution. Despite this restricted framework, the selected balances play a key role in shaping the flow of extratropical cyclones, governing the atmospheric dynamics in some regions of the system.

\subsection{Hydrostatic Balance} 

The MEPS model is based on the HARMONIE–AROME system, which employs a non-hydrostatic dynamical core that solves the fully compressible Euler equations \citep{Bengtsson2017,Gleeson2024}, including the complete vertical momentum equation \citep{Laprise1992}. As a result, MEPS does not enforce hydrostatic balance (HB), defined as the equilibrium between the downward gravitational force and the upward vertical pressure gradient force. Deviations from HB should not be regarded as unphysical, as the atmosphere is often imbalanced. Such imbalances play a fundamental role in driving adjustment processes and temporal evolution, and are therefore an essential component of any NWP system.

In particular, significant departures from HB are expected in dynamically active situations, such as severe windstorms and extratropical cyclones with sharp frontal structures. Consequently, in this study we do not assess the physical realism of Bris based on its ability to establish a hydrostatically balanced atmosphere. Instead, we evaluate its internal consistency by quantifying deviations from HB, analysing their spatial distribution, and assessing whether these imbalances are coherent with those produced by MEPS and with the structure of the windstorm Poly in the analysis and the NWP forecasts.

The datasets from Bris, MEPS, and the corresponding analyses are provided on isobaric pressure coordinates, with pressure as the vertical coordinate. In this framework, the hydrostatic equation is:
\begin{equation}
\left( \frac{\partial \Phi}{\partial p} \right)_{\mathrm{HB}} = -\frac{R T_v}{p},
\label{eq:hydrostatic_equation}
\end{equation}
where \( R = 287\,\mathrm{J\,kg^{-1}\,K^{-1}} \) is the specific gas constant for dry air, and \( T_v = T(1 + 0.61q) \) is the virtual temperature.

Following \eqref{eq:hydrostatic_equation}, we compute the theoretical geopotential thickness \( d\Phi_{\mathrm{HB}} \) between two adjacent pressure levels \( p_i \) and \( p_{i+1} \) (with \( p_{i+1} < p_i \)) as:
\begin{equation}
d\Phi_{\mathrm{HB}} = -R \cdot \frac{\frac{1}{2}\bigl[T_v(p_i) + T_v(p_{i+1})\bigr]}{\frac{1}{2}(p_i + p_{i+1})} \cdot (p_{i+1} - p_i).
\label{eq:dPhi_HB}
\end{equation}
The geopotential thickness $d\Phi$ between these consecutive pressure levels is simply defined as the difference between the corresponding predicted geopotential values. 

We then define a dimensionless deviation metric at the layer midpoint pressure $p_{i+1/2}$ as:
\begin{equation}
\delta_{\mathrm{HB}}(p_{i+1/2}) = \frac{d\Phi - d\Phi_{\mathrm{HB}}}{d\Phi}.
\label{eq:delta_HB}
\end{equation}
This metric quantifies the relative departure of the predicted geopotential thickness from the one expected under HB, for each layer and grid point.

\subsection{Geostrophic Balance}

When friction, local time derivatives, and advective terms are neglected, the Navier--Stokes momentum equation reduces to a stationary balance between the Coriolis force and the geopotential gradient force \citep{CushmanRoisin2006}. Under this assumption, the geostrophic wind is given by
\begin{equation}
\mathbf{v}_g = \frac{1}{f}\,\hat{\mathbf{k}} \times \nabla_p \Phi ,
\label{eq:geostrophic_wind}
\end{equation}
where \(f\) is the Coriolis parameter and \(\nabla_p\) denotes the horizontal gradient on isobaric surfaces. Neither MEPS nor the analysis fields are expected to satisfy Eq.~\eqref{eq:geostrophic_wind} exactly, as the neglected terms are generally non-zero and solved explicitly in the Harmonie-Arome dynamical core \citep{Bengtsson2017,Laprise1992}. Departures from geostrophic balance reflect the presence of ageostrophic motions and the imbalance results in local accelerations. Nevertheless, in some specific atmospheric regions and over large spatial scales, geostrophic balance often holds in a first order approximation.

To identify such regions, in this study, ageostrophic winds are computed for MEPS, Bris and the analysis fields only at grid points within the \(443 \times 443\) subdomain where the Rossby number,
\begin{equation}
\mathrm{Ro} = \frac{\left\| \left( \mathbf{v} \cdot \nabla_p \right) \mathbf{v} \right\|}{f \left\| \mathbf{v} \right\|},
\label{eq:Ro_threshold}
\end{equation}
falls below a prescribed threshold. We used $\mathrm{Ro} < 0.1$ for the results in Fig.~\ref{fig:figure6}; a threshold which ensures enough grid points and robust statistics; alternative thresholds (0.15, 0.2, 0.25, and 0.3) were also tested with consistent results. Small values of $\mathrm{Ro}$ indicate that horizontal advective terms are relatively small compared to the Coriolis force, and that geostrophic balance is likely to be a good approximation. Additionally, this criterion restricts the analysis primarily to regions outside the Poly cyclone’s LPC, where we observed vertical advection to be relatively weak and large-scale rotational flow to dominate, which makes the assumption of near-dynamical equilibrium (i.e., negligible local time accelerations) more appropriate. 

To obtain the ageostrophic wind components, we then compute the Euclidean norm of the difference between the predicted wind and the geostrophic wind:
\begin{equation}
ws_{ag} = \left\| \mathbf{v} - \mathbf{v}_g \right\|
\label{eq:geostrophic_imbalance}
\end{equation}
. This metric is used to quantify local departures from geostrophic equilibrium, in the locations where geostrophic balance is expected to approximately hold ($\mathrm{Ro} < 0.1$).

\subsection{Gradient Wind Balance}

The chosen Rossby number threshold for the geostrophic balance analysis primarily selects grid points outside the cyclone’s LPC, where inertial effects are relatively weak. This leaves us with limited knowledge about whether dynamical balances are maintained within the core of the system. Near the centre of the extratropical cyclone, the strong rotational flow gives rise to a substantial centrifugal force, which becomes non-negligible in the momentum balance. In this region, the balance involves three terms: the centrifugal force, the Coriolis force, and the pressure-gradient force; an equilibrium commonly referred to as the gradient wind balance and expressed as:
\begin{equation}
\quad \quad \frac{u_{\theta}^2}{r} + f u_{\theta} = \frac{\partial \Phi}{\partial r},
\label{eq:gradient_wind_balance}
\end{equation}
where $r$ denotes the radial distance from the LPC of the cyclone, and $u_{\theta}$ the tangential (azimuthal) velocity expressed in a cylindrical coordinate framework \citep{CushmanRoisin2006}.

The gradient wind balance is not required to hold exactly in either the MEPS forecasts or the corresponding analysis fields. Indeed, this balance is obtained by neglecting some terms in the momentum equation expressed in cylindrical coordinates \citep{CushmanRoisin2006}. In the dynamical core of Harmonie-Arome underlying the MEPS system, the omitted terms do not vanish and can introduce imbalances, leading to non-zero local wind accelerations \citep{Bengtsson2017,Laprise1992}. Furthermore, the ideal gradient wind balance arises in a steady-state flow, where the LPC of the cyclone is nearly stationary and its intensity does not change over time. However, in a moving and evolving extratropical cyclone like Poly this condition is not properly satisfied. Therefore, similar to the HB study, rather than evaluating the consistency of the Bris output by looking at the magnitude of the gradient wind balance in its predictions, we decided to evaluate if the deviations from gradient wind balance are reasonable, from our physical expectations on the cyclone dynamics as well as from a comparison with the deviations seen in the NWP model and the analyses. 

Contrary to the other balances, the examination of the gradient wind balance requires a key methodological change. Rather than using the raw geopotential fields predicted by the models, we use fitted geopotential fields. This choice is motivated by two main considerations. First, from a practical standpoint, expressing the geopotential as an explicit function of the radial coordinate \( r \) enables a straightforward evaluation of the radial gradient term \( \partial \Phi / \partial r \). Second, from a methodological perspective, the geostrophic balance analysis revealed the presence of fine-scale noise in the DDM output, which makes the computation of spatial derivatives over short distances unreliable. By fitting the geopotential field, this noise is effectively filtered out, allowing us to assess whether the gradient wind balance is satisfied at larger, dynamically relevant spatial scales.

The procedure used to fit the geopotential fields and to derive the tangential component of the gradient wind, $u_{\theta}$, consists of four main steps and follows the methodology proposed by \citet{GarciaFranco2021}. The four steps are outlined here, and supplementary Fig. S2 provides a schematic overview to facilitate understanding.
\begin{enumerate}[topsep=0pt]
    \item The cyclone LPC is identified as the minimum of the geopotential field within a $443 \times 443$ grid-point subdomain. This identification is supported by a visual inspection to verify that the detected geopotential minimum is reasonably aligned with the storm’s LPC.
    \item The geopotential field is recentered on the identified LPC and extracted within a radius of 125~km, which effectively corresponds to evaluating the model fields in a storm-relative reference frame.
    \item The recentered geopotential field is fitted by a function $\Phi_{\text{fit}}(r,\theta)$, obtained from a truncated radial--azimuthal polynomial expansion with maximum orders $N,M \leq 5$:
    \begin{equation}
    \Phi_{\text{fit}}(r, \theta) = 
    \sum_{n=0}^{N} \sum_{m=0}^{M} 
    \left[ a_{nm} \, r^n \cos(m\theta) + b_{nm} \, r^n \sin(m\theta) \right],
    \label{Phi_fit}
    \end{equation}
    where $\theta$ denotes the azimuthal angle, and $a_{nm}$ and $b_{nm}$ are the fitted coefficients. To ensure that the results were robust to the choice of truncation order, several $(N,M)$ pairs were tested (e.g., $N,M = 1,2,3,4,5,6$). The choice $N = M = 5$ provided the most accurate visual representation of the geopotential field.
    \item Owing to the explicit radial dependence of $\Phi_{\text{fit}}$, the radial derivative $\partial \Phi / \partial r$ can be directly computed. Substituting this expression into Eq.~\eqref{eq:gradient_wind_balance} yields a second-order equation for the azimuthal velocity $u_{\theta}$, whose solution provides the gradient wind speed (namely, $ws_{grad}$).
\end{enumerate}

\subsection{Continuity Equation}

A fundamental kinematic constraint linking the horizontal and vertical components of the flow is provided by the continuity equation. Assessing whether this constraint is satisfied offers a direct and physically meaningful test of the internal consistency between the model-resolved wind divergence and vertical motion. The continuity equation in pressure coordinates reads:
\begin{equation}
\frac{\partial \omega}{\partial p} = - \nabla_p \cdot \mathbf{v},
\label{eq:continuity_equation}
\end{equation}
which implies an expected anticorrelation between the vertical gradient of the pressure velocity,
$\partial \omega / \partial p$, and the horizontal divergence of the wind field,
$\nabla_p \cdot \mathbf{v}$. Nevertheless, departures from this idealized 
relationship are commonly observed even in physics-based models, owing to factors such as 
non-hydrostatic motions, diabatic processes, finite numerical resolution, and the 
discretization approximations involved in computing the relevant fields.

To study the continuity equation, we first compute the difference in vertical velocity between two consecutive pressure levels:
\begin{equation}
\frac{\partial \omega}{\partial p}\,dp = \omega(p_{i+1}) - \omega(p_i)
\label{eq::dwdp_dp}
\end{equation}
where \(\omega = \partial p/\partial t\) is the vertical velocity in pressure coordinates. The horizontal wind divergence over the same pressure layer is then obtained as the average horizontal wind divergence between the upper and lower levels, integrated over the layer thickness to ensure consistent units with the vertical velocity difference:
\begin{equation}
\nabla_p \cdot \mathbf{v}\, dp =
\frac{1}{2}\Big(
\frac{\partial u}{\partial x}(p_{i+1}) + \frac{\partial v}{\partial y}(p_{i+1})
+ \frac{\partial u}{\partial x}(p_i) + \frac{\partial v}{\partial y}(p_i)
\Big)\,(p_{i+1}-p_i)
\label{eq::div_v}
\end{equation}

We then perform a qualitative inspection to assess the degree of anticorrelation between the two fields in Bris, MEPS outputs, and the analysis fields. 

\section{Results}

\subsection{A first general assessment}

\begin{figure}[h!]
    \centering
    \includegraphics[width=0.9\textwidth]{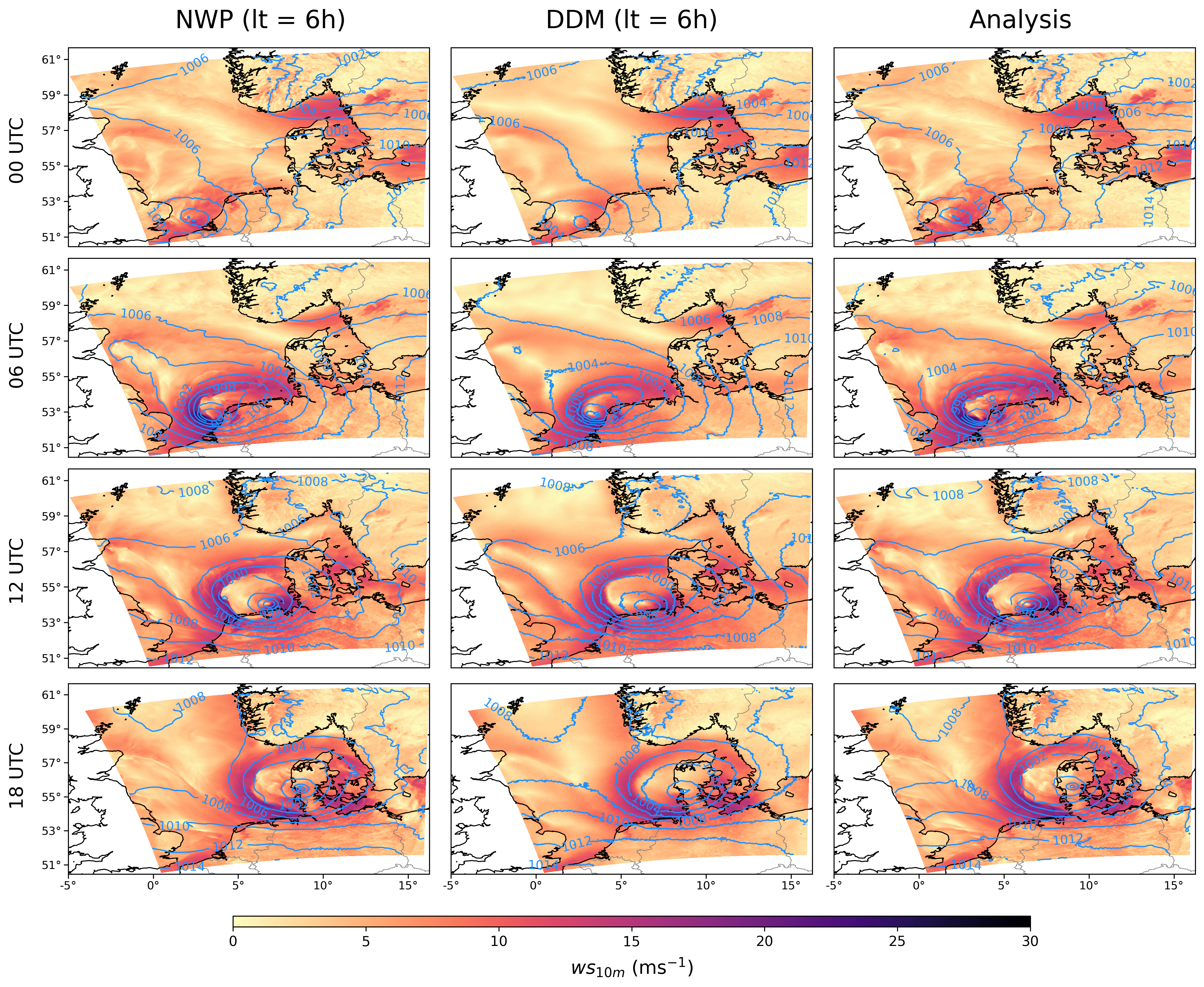}
    \caption{\small Maps of 10m wind speed forecasts (color) and mean sea-level pressure (blue contours, in hPa) at +6h lead time for the MEPS (NWP; left) and Bris (DDM; centre) model, together with the corresponding analysis fields (right). Fields are provided for four different times during the occurrence of windstorm Poly: 00, 06, 12, 18UTC on 5 July 2023 (from top to bottom).}
    \label{fig:figure1}
\end{figure}

\begin{figure}[h!]
    \centering
    \includegraphics[width=0.9\textwidth]{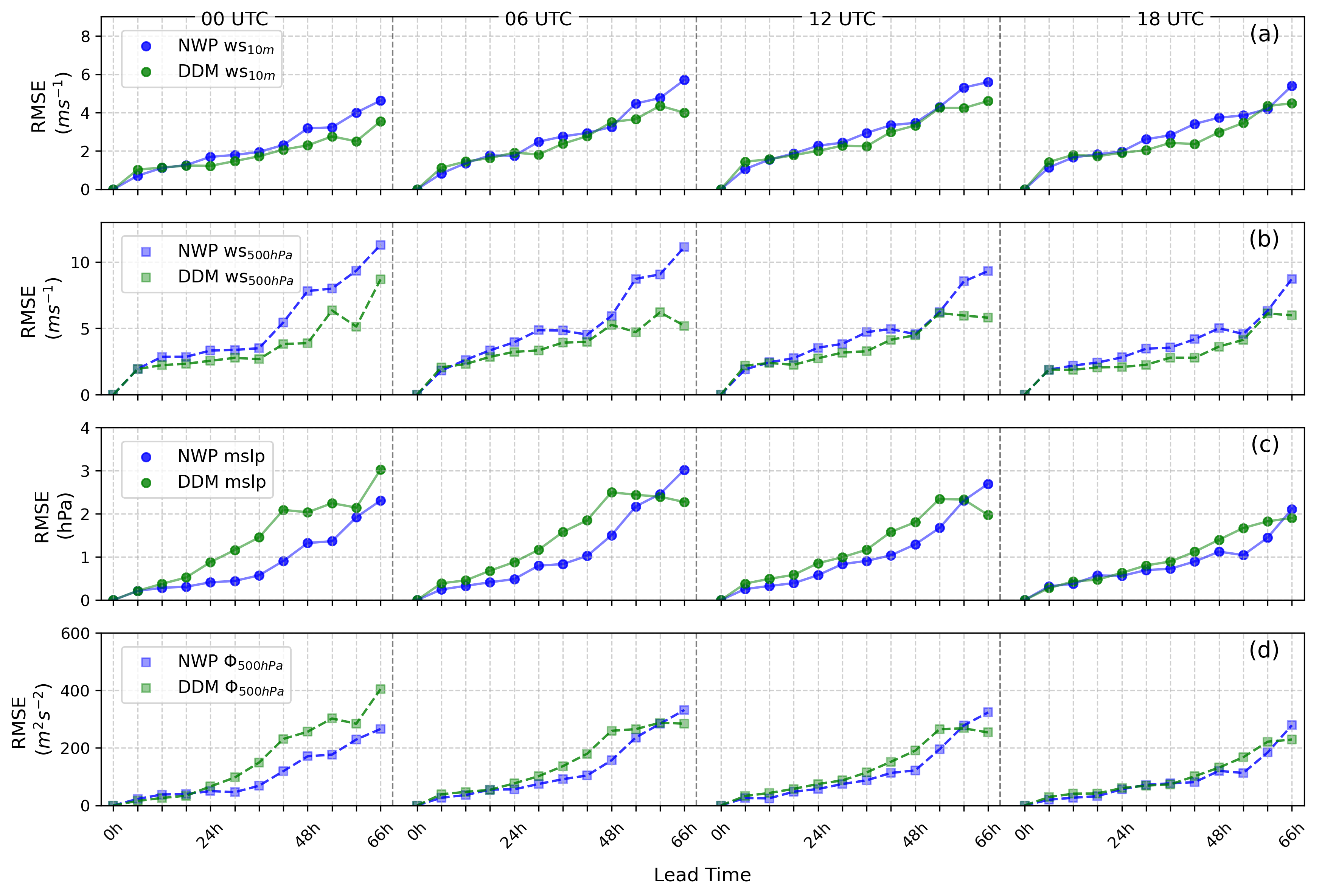}
    \caption{\small RMSE of (a) 10m wind speed, (b) 500hPa wind speed, (c) mean sea-level pressure, (d) 500hPa geopotential in the selected 443x443 domain (Fig. S1). Values are given as a function of lead time for four times (00, 06, 12, 18 UTC) on 5 July 2023, during the occurrence of windstorm Poly.}
    \label{fig:figure2}
\end{figure}

From Fig.~\ref{fig:figure1}, we find that, for a +6h lead time, Bris captures the large-scale dynamics of the storm well and performs overall comparably to MEPS. Bris successfully reproduces the strong near-surface wind speeds associated with storm Poly, accurately identifying the regions of maximum intensity. Additionally, the predicted locations of the cyclone’s LPC closely match those in the analyses and are of similar accuracy to the NWP forecasts. For instance, at 06 UTC, the LPC position predicted by Bris (that we here regard as the location of minimum mean sea-level pressure) is displaced by approximately 17 km from the centre identified in the analysis field, compared with a displacement of about 14 km for MEPS. Across different forecast initialization times and lead times, several cases are identified in which Bris does locate the storm centre even more accurately than MEPS, indicating that its skill in representing the storm track is, at times, comparable to or better than that of the NWP model (not shown).

However, the DDM tends to generate excessively smoothed meteorological fields already from lead time 6h. This tendency is a well-known limitation of DDMs trained using an MSE loss function \citep[e.g.,][]{Lam2023,Lang2024}, and has already been reported for Bris by \citet{Nipen2024}. We also observed (\textit{not shown}) that this large-scale smoothing becomes progressively more pronounced with increasing lead time, leading to an increasingly pronounced underestimation of extreme features, like the highest wind speeds appearing near the cyclone core. In particular, during the mature phase of the storm at 06 UTC and for +6h forecasts, the maximum wind speed predicted by the DDM is 23.5~m~s$^{-1}$, lower than the values predicted by the NWP model and found in the analysis (26.1 and 26.4~m~s$^{-1}$, respectively).

Despite this underestimation of extremes, Bris demonstrates robust performance for the actual metric Bris was trained to minimize (MSE). The DDM exhibits lower RMSE values than the NWP model both near the surface and in the upper atmosphere (see Fig.~\ref{fig:figure2}a-b). Therefore, although Poly represents a rare event likely underrepresented in the 3.3 years of high-resolution training data, Bris still outperforms the MEPS control member in terms of wind speed prediction accuracy. In contrast, for mean sea-level pressure and geopotential at 500hPa, the MEPS control member generally shows better performance with lower RMSE, except at longer lead times of 60–66h, where Bris performs equally well or even better (Fig.~\ref{fig:figure2}c-d). In general, when considering RMSE alone, we noticed small differences between Bris and the MEPS control member, and that for some variables Bris even surpasses the MEPS control member, in agreement with \citet{Nipen2024}.

\subsection{Hydrostatic balance}

\begin{figure}[h!]
    \centering
    \includegraphics[width=1\textwidth]{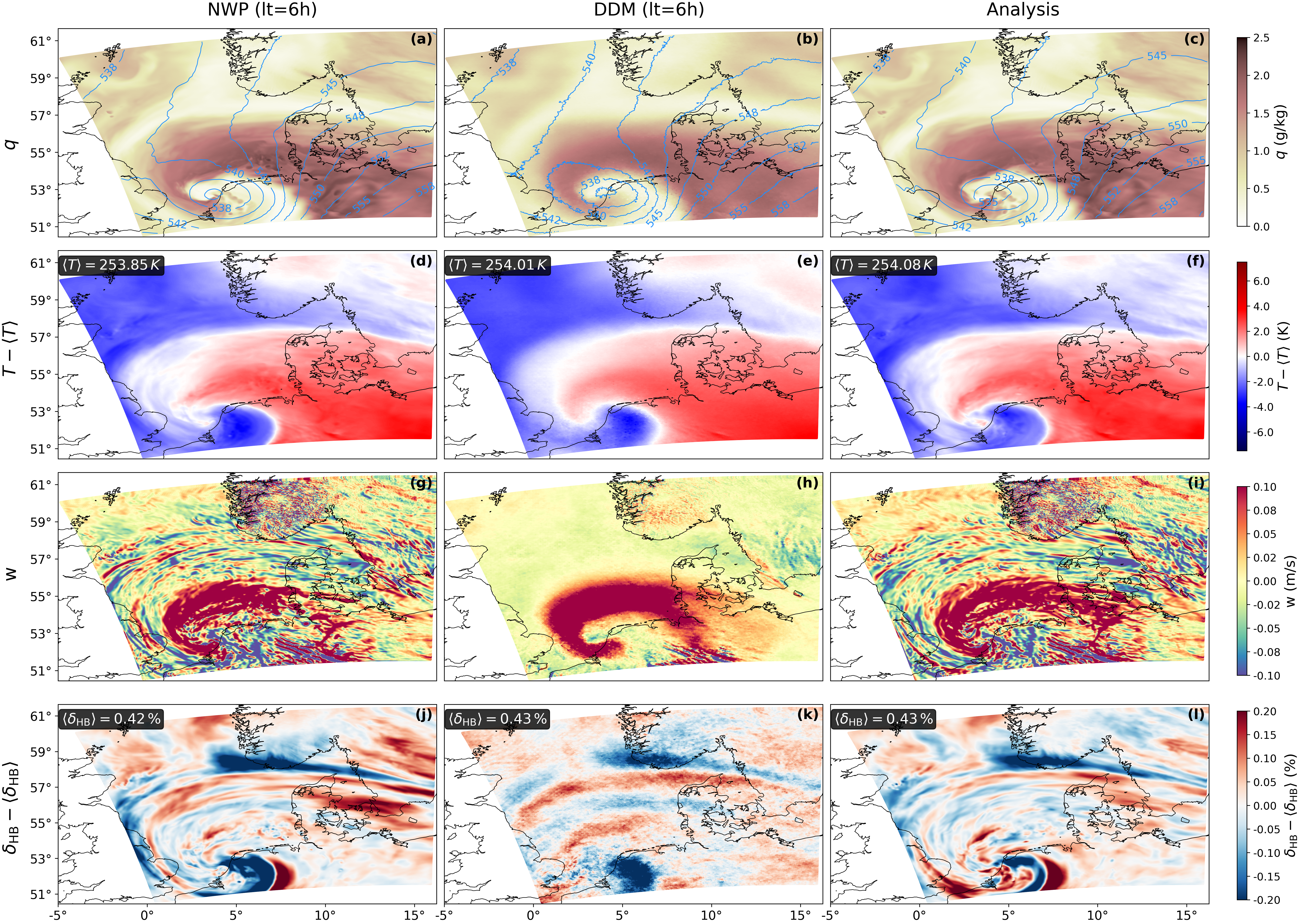}
    \caption{\small MEPS (NWP; left) and Bris (DDM; centre) +6h forecasts, along with the corresponding analysis fields (right) at 06 UTC on 5 July 2023 (windstorm Poly at its mature stage), of: (a–c) 500 hPa specific humidity (shading) and geopotential (blue contours, in $10^2\,\mathrm{m^2\,s^{-2}}$); (d–f) 500hPa temperature anomalies; (g–i) 500hPa vertical velocity; and (j–l) hydrostatic balance anomalies between 500hPa and 400hPa. Anomalies are computed relative to the domain-mean values of each field over the 443x443 domain shown in the upper left corner of panels (d)-(f) and (j)-(l)}
    \label{fig:figure3}
\end{figure}

Fig.~\ref{fig:figure3}a--f show that Bris reproduces the specific humidity and the temperature fields in close agreement with the analysis, despite some spatial smoothing. In contrast, the DDM-predicted vertical velocity field exhibits substantial discrepancies relative to both the NWP output and the analysis, which are in much closer agreement with each other (Fig.~\ref{fig:figure3}g--i). In the DDM forecast, a continuous spiral band of positive vertical velocity emerges in the northern sector of the cyclone’s LPC, a key characteristic of North Atlantic extratropical cyclones \citep[e.g.,][]{Bjerknes1919, BjerknesSolberg1922, Laurila2021}. This large-scale ascent reflects the convergence of warm, moist air advected from the southeast with colder, denser air descending from the northwest.

In this respect, the DDM provides a physically interpretable representation of the large-scale upward motion, but fails to capture the finer mesoscale convective structures evident in the NWP output and analysis. These fields display a markedly richer pattern of localized updrafts and downdrafts, which are largely absent in the DDM predictions (Fig.~\ref{fig:figure3}g--i). The resulting smoothness of the vertical velocity signal in Bris can be attributed to the model’s training configuration, where vertical velocity was intentionally assigned a lower weight in the loss function \citep{Lang2024,Nipen2024}. Moreover, the limited occurrence of severe windstorms within the high-resolution limited domain might also have played a role in the model’s difficulty to learn complex and localized convective features associated with Poly. Consistently, the intricate pattern of updrafts and downdrafts over Norway’s mountainous regions, clearly visible in both the NWP output and the analysis, is strongly underrepresented in the DDM forecast (Fig.~\ref{fig:figure3}g--i).

We next examine the HB anomaly fields (Fig.~\ref{fig:figure3}j--l). The domain-averaged deviations from HB, shown in the black boxes, are positive and remarkably similar across NWP, DDM, and analysis datasets, indicating that the geopotential increases more rapidly with altitude than expected at 500–400 hPa from the hydrostatic assumption. This systematic positive deviation may partly stem from the simplified mid-layer temperature used in computing $d\Phi_{\text{HB}}$, given the coarse vertical resolution of just 12 pressure levels. Alternatively, it could reflect physically meaningful upward motion associated with intense convection and frontal interactions at lower pressure levels. Overall, the DDM reproduces the mean deviations well,  suggesting no major violation of the HB on average.

More insight arises from comparing the spatial distribution of HB anomalies with the expected dynamics of the extratropical cyclone. In Fig.~\ref{fig:figure3}j--l, during Poly’s mature stage, alternating negative (blue) and positive (red) regions appear over the domain. These patterns likely reflect compression and expansion of geopotential isosurfaces driven by strong vertical accelerations and decelerations of the wind, with consequent localized disruptions from HB. Such pronounced blue and red regions are likely to form near the interfaces between warm and cold fronts in the southern part of the cyclone, where intense gusts of warm air collide with opposing cold air masses, producing substantial vertical activity. This physical behavior seems clearly captured in the analysis field (Fig.~\ref{fig:figure3}l), which shows three strongly colored regions in the southeast sector of the storm's core, corresponding to the inferred locations of the three fronts.

The Bris model successfully captures the dominant blue region in the southern sector of the LPC (Fig.~\ref{fig:figure3}k), which could be explained by the strong subsidence of cold air at lower levels that depresses the geopotential surfaces and reduces the vertical geopotential gradient in the overlying atmosphere; making $\delta\Phi_{\text{pred}} < \delta\Phi_{\text{HB}}$ at the 500–400 hPa mid-level. However, at the same pressure level, Bris fails to reproduce the adjacent red anomalies, which would be indicative of warm air ascending over colder air in the lower troposphere, thereby enhancing the vertical geopotential gradient in the overlying 500–400 hPa layer. Instead, these red anomalies are better represented in the NWP prediction (Fig.~\ref{fig:figure3}j), although the upward advection of warm air into the cyclone’s core remains underestimated with respect to the analysis field.

A further aspect deserving attention in the HB anomaly fields (Fig.~\ref{fig:figure3}j--l) is the appearance of filamentary structures that align with and follow the spiral circulation of the cyclone. These features can be interpreted as disruptions of HB being advected along the flow, since the azimuthal wind component is typically much stronger than the radial one. Consequently, HB anomalies are transported along spiral trajectories, along with other dynamical properties such as moisture, mesovortices, and potential vorticity \citep[e.g.,][]{Rozoff2006,Wang2008}. These filamentary patterns are also reasonably well reproduced by the Bris output, although the NWP fields remain evidently closer to the filament shapes appearing in the analysis.

\subsection{Geostrophic balance}
\label{geo_bal_section}

\begin{figure}[h!]
    \centering
    \includegraphics[width=1.0\textwidth]{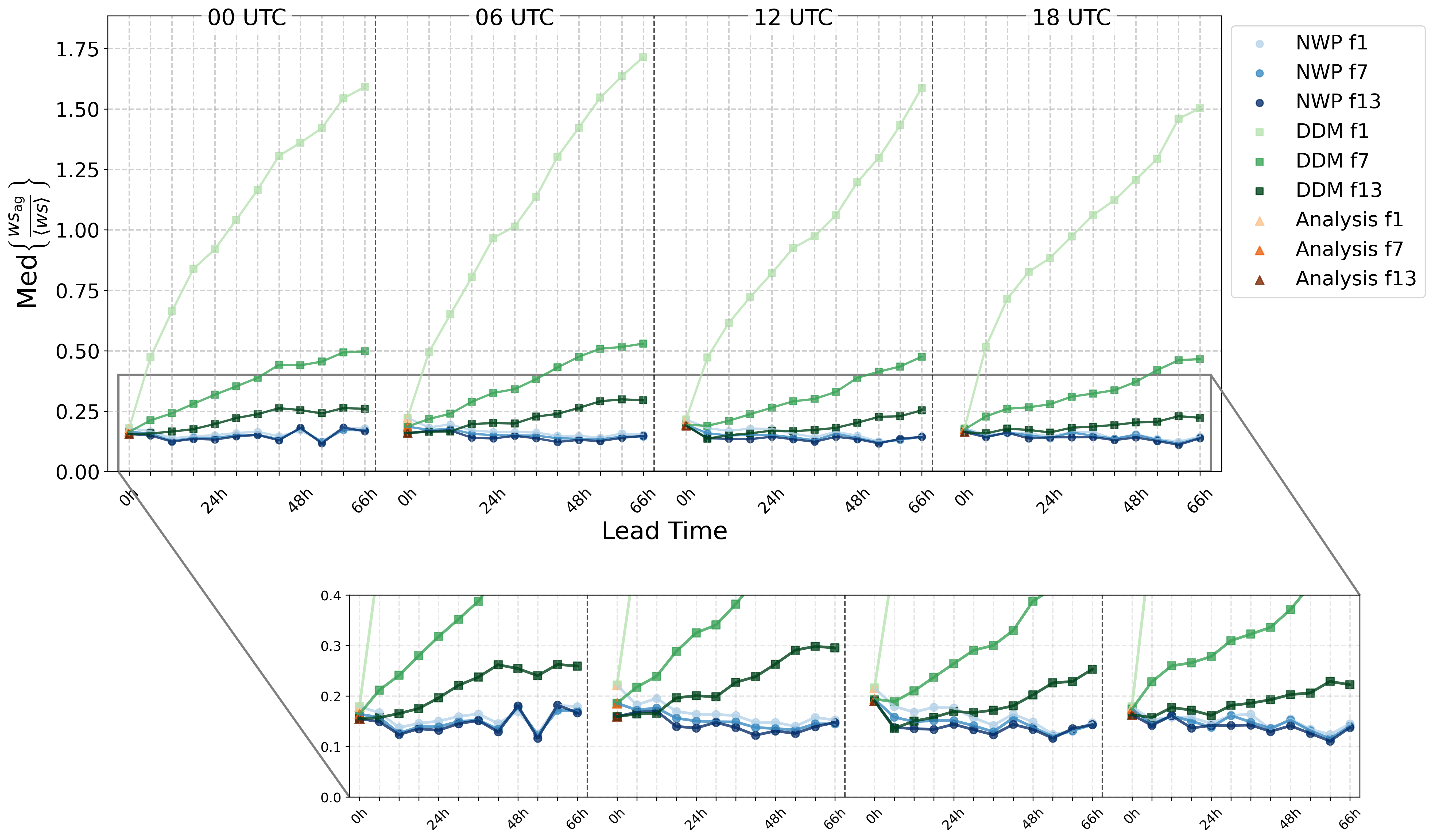}
    \caption{\small Spatial median normalized ageostrophic wind speeds in MEPS (NWP) forecast, Bris (DDM) predictions, and analysis fields, shown as functions of lead time and computed at locations where Ro<0.1 within the selected 443×443 domain. Results refer to 500hPa winds at four times (00, 06, 12, 18 UTC) on 5 July 2023, when windstorm Poly moved over the domain. Ageostrophic components are normalized by the mean wind speed of the distribution and are presented for three cases: f1 (no smoothing), f7 and f13 (winds computed after smoothing; see section~\ref{sec3}).}
    \label{fig:figure4}
\end{figure}

As described in the methodology, physical balances are first analyzed using the raw model outputs and subsequently re-examined after applying artificial smoothing to the fields. Fig.~\ref{fig:figure4} shows the median deviations from geostrophic balance as a function of lead time, at four different phases of the storm Poly. Focusing initially on the raw outputs (i.e., the unsmoothed fields, denoted by the f1 label), we observe that the DDM outputs exhibit systematically higher curves compared to both the NWP model and the analysis, which are in close agreement with each other. These unrealistically high and inconsistent ageostrophic wind magnitudes in the DDM highlight the model's difficulty in producing geopotential fields and horizontal winds that closely adhere to geostrophic balance.

\begin{figure}[h!]
    \centering   
    \includegraphics[width=0.9\textwidth]{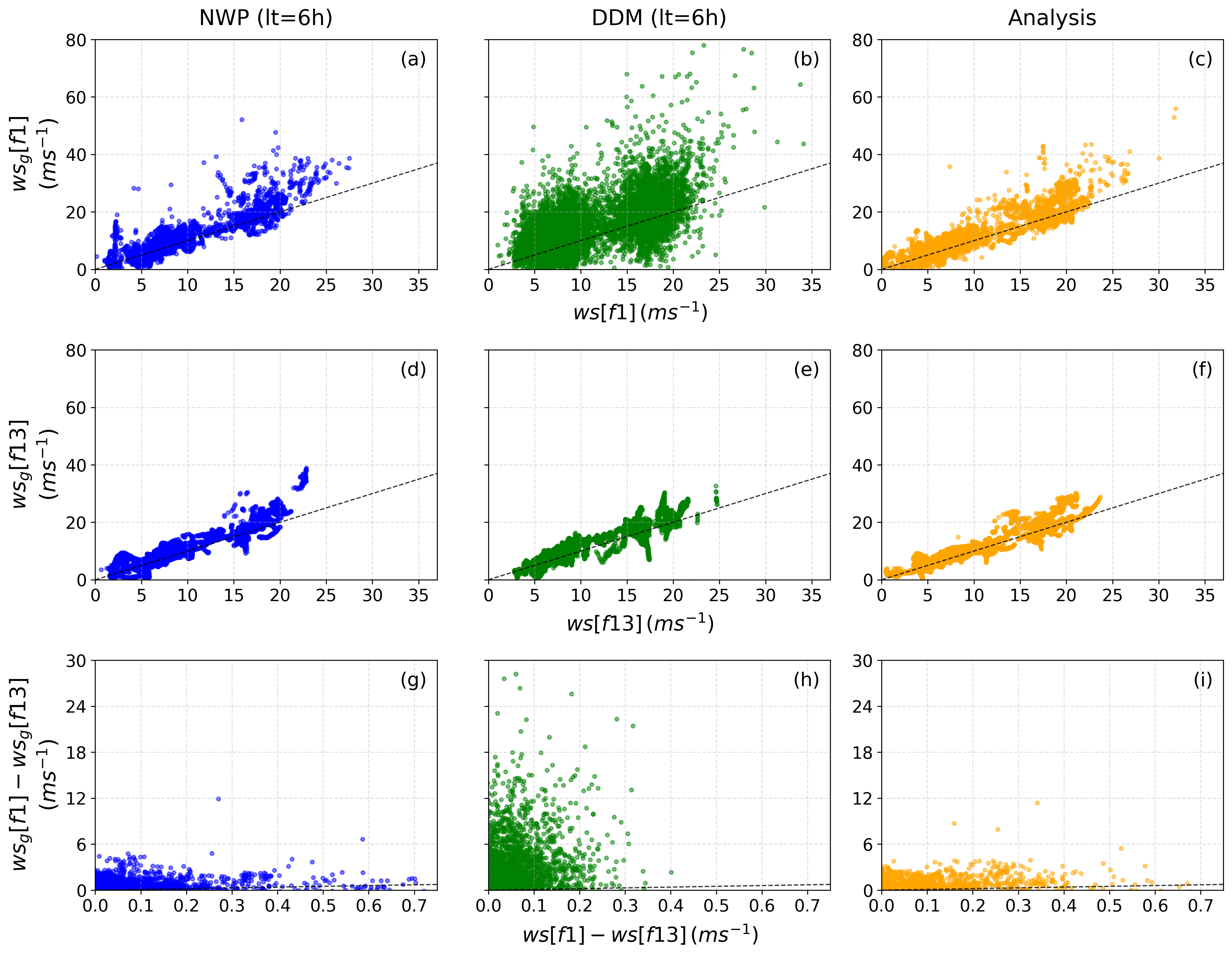}  
    \caption{\small Scatter plots of 500hPa predicted wind speed versus geostrophic wind speed for (a–c) unsmoothed fields (f1), (d–f) winds computed after smoothing (f13), and (g–i) noise-to-noise comparisons. Values are given for MEPS (NWP; left), Bris (DDM; centre) +6h forecasts, and the corresponding analysis (right) at 06 UTC on 5 July 2023, during the occurrence of windstorm Poly, considering only grid points with Ro<0.1 within the selected 443 $\times$ 443 domain. The dashed black line (y = x) indicates perfect agreement.}
    \label{fig:figure5}
\end{figure}

The results appear largely independent of the specific valid time considered, suggesting that they reflect an intrinsic property of the model forecasts rather than a response to the particular event or atmospheric state examined.
For all times (00, 06, 12 and 18UTC), the DDM shows a systematic increase in median ageostrophic winds with lead time. In contrast, the NWP model maintains a comparatively smaller and stable median imbalance, which could be expected as its forecasts remain always constrained by physical equations, regardless of the lead time considered. 

Another key result shown in Fig.~\ref{fig:figure4} is that progressively smoothing the fields from the DDM outputs (lines f7 and f13) allows to get fields closer to geostrophic balance, with median ageostrophic components approaching those observed in the NWP forecasts and analyses. The same results are confirmed in Fig.~\ref{fig:figure5}a--b, which present scatterplots of geostrophic wind speed versus predicted wind speed. Particularly, in Fig.~\ref{fig:figure5}a, raw DDM predicted winds show poor agreement with geostrophic winds, whereas raw NWP and analysis predictions display better agreement and are relatively similar to each other, with deviations tending toward subgeostrophic winds for stronger gradients. After the f13 smoothing (Fig.~\ref{fig:figure5}b), the DDM scatterplot becomes more consistent with what is observed in the NWP outputs and the analyses.

Finally, we extend the analysis by including horizontal wind advection alongside the two terms defining geostrophic balance, and we examine the equilibrium arising from the three dominant forces in the momentum equation: advection, the pressure gradient, and the Coriolis force. In this case, rather than using the Rossby number as a threshold to identify regions where equilibrium is expected, we impose a constraint on wind convergence. This analysis yields results that are highly consistent with those obtained and presented for the geostrophic balance. The methodology and results related to this extended physical relationship are described in detail in the MSc thesis \citep{Pasquini2025}, which we refer to for completeness and clarity.

\subsection{Gradient wind balance}

\begin{figure}[h!]
    \centering
    \includegraphics[width=0.9\textwidth]{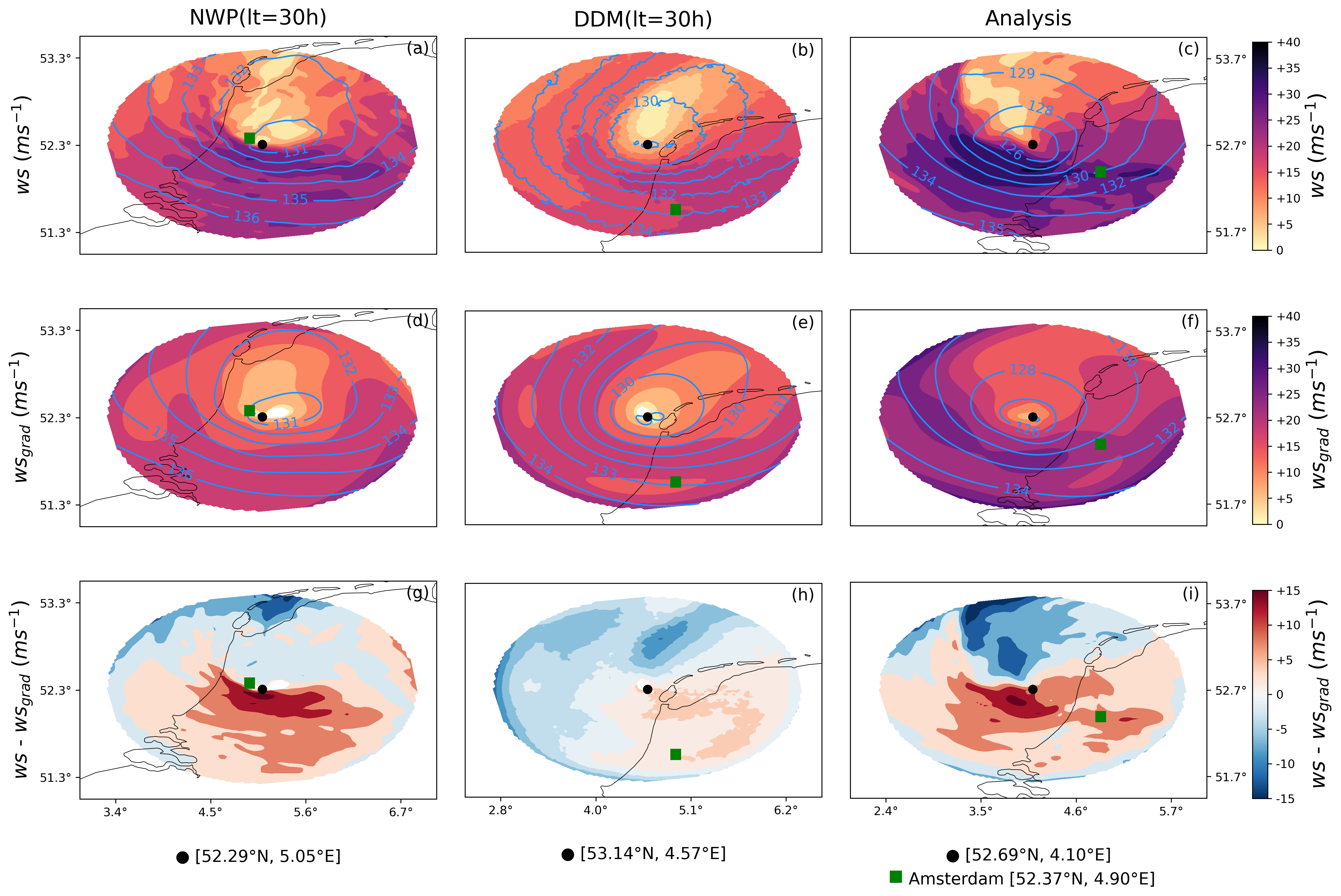}
    \caption{\small (a–c) Wind speed and geopotential fields at 850hPa (blue contours, in $10^2$\,$m^2s^{-2}$); (d–f) corresponding gradient wind fields computed from the fitted geopotential (blue contours, in $10^2$\,$m^2s^{-2}$); (g–i) difference between wind speed and gradient wind. Plots are presented in a cyclone-coherent reference frame, within a 125 km radius from the predicted LPC position (black dot) of storm Poly. Columns show MEPS (NWP; left) and Bris (DDM; centre) +30h forecasts and the corresponding analysis field (right). All fields are valid at 06 UTC on 5 July 2023, when Poly reached its mature stage and approached the Dutch coast.}
    \label{fig:figure6}
\end{figure}

Fig.~\ref{fig:figure6}a–c indicate that the spatial distribution of wind speeds in the +30h NWP forecasts closely resembles the analysis, whereas the DDM systematically underestimates wind speeds. Although slightly displaced and weaker, the strong winds ($>25~\mathrm{ms^{-1}}$) affecting the southern sector of storm Poly at 06 UTC—impacting Amsterdam and other parts of North Holland—are well captured by the NWP forecast, while they are strongly reduced (by at least $10~\mathrm{ms^{-1}}$) in the DDM field. In addition, Fig.~\ref{fig:figure6}b highlights much noisier isogeopotential contours in the DDM than in the corresponding NWP and analysis fields.

Despite the erroneous representation of wind speeds within the cyclone core, Bris performs comparably to—and in some cases better than—MEPS in locating the LPC of storm Poly and predicting the minimum geopotential. Moreover, the associated departures from gradient wind remain physically realistic. In particular, Fig.~\ref{fig:figure6}h demonstrates that the DDM forecast does not exhibit larger departures from gradient-wind balance; instead, deviations are systematically smaller and largely confined within $-5$ to $+5~\mathrm{ms^{-1}}$. By contrast, both the NWP forecast and the analysis feature deviations that can reach values up to three times larger. This apparently contrasting behavior relative to that observed for the geostrophic balance (Fig.~\ref{fig:figure4}), where the DDM produces unrealistically large ageostrophic winds compared to the numerical predictions, is discussed and physically interpreted in Section~\ref{sec:discussion}.

Importantly, non-negligible departures from gradient-wind balance are observed even in physics-based predictions (Fig.~\ref{fig:figure6}g–i), ranging from approximately $-15$ to $+15~\mathrm{ms^{-1}}$ in regions where observed wind speeds locally exceed $40~\mathrm{ms^{-1}}$. These deviations should not be interpreted as errors or inconsistencies. While the gradient-wind approximation provides a useful first-order description of the flow, it does not fully capture the wind pattern near the extratropical cyclone core. Such departures are physically expected and can arise from a variety of processes, including locally intense convective activity, along-flow tangential acceleration, and—particularly relevant in this case—the superposition of the cyclone’s translational motion onto the gradient-wind field \citep{Loridan2013}.

\begin{figure}[h!]
    \centering
    \includegraphics[width=0.7\textwidth]{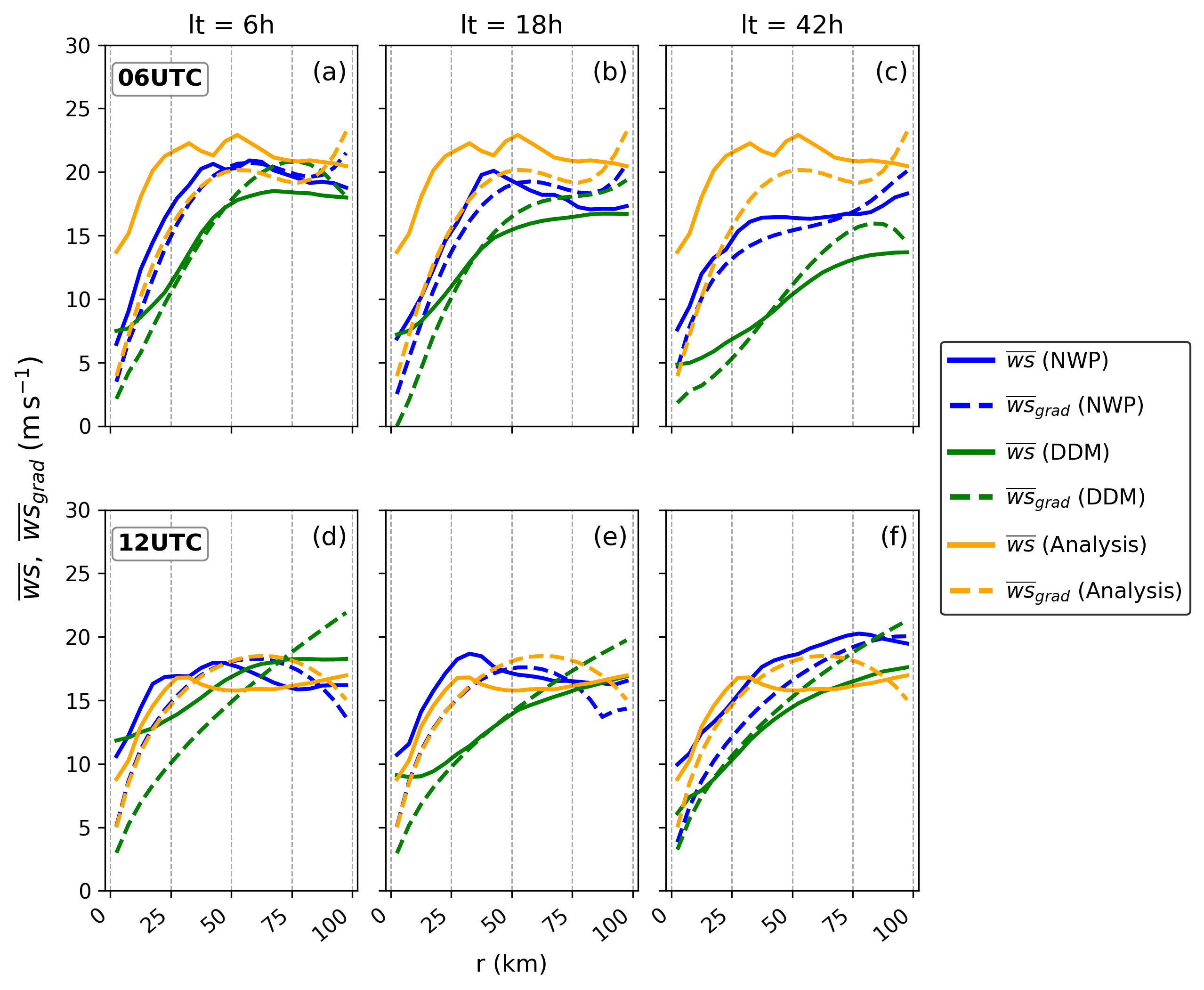}
    \caption{\small Azimuthally averaged wind speed ($\overline{ws}$; solid lines) and gradient wind ($\overline{ws}_{grad}$; dashed lines) at 850hPa for (a--c) 06~UTC and (d--f) 12~UTC, shown as a function of radial distance $r$ (km) from the predicted location of the cyclone LPC. In all panels, blue, green, and orange denote the MEPS (NWP) model, Bris (DDM), and Analysis, respectively. NWP and DDM values correspond to forecasts with lead times +6h (left), +18h (centre), and +42h (right), whereas the Analysis is shown identically in all panelsof each row for the same time. Averages are computed over 5km-wide rings centered on the cyclone.}
    \label{fig:figure7}
\end{figure}

The examination of the spatial distribution of gradient-wind deviations further highlights the closer agreement between the NWP forecast and the analysis. Nevertheless, Bris appears to capture, at least partially, the large-scale pattern of these deviations. For instance, in the +36 h forecasts shown in Fig.~\ref{fig:figure6}g–i, structured deviations from gradient wind are evident in the northern and southern sectors of the storm’s LPC across all datasets (NWP, DDM, and analysis). These deviations create a pronounced asymmetry in the $ws - ws_{grad}$ field, with a negative anomaly in the northern portion of Poly’s LPC and a positive anomaly in the southern portion. Physically, this asymmetry is consistent with the cyclone’s eastward motion: in the southern sector, the rotational gradient wind reinforces the eastward translation, while in the northern sector it acts against it. This physically expected asymmetry is also present in the DDM forecasts, although systematically underestimated. 

Fig.~\ref{fig:figure7} shows another key result: in the DDM forecasts, the radial profiles of azimuthally averaged predicted wind speeds and the corresponding gradient winds weaken consistently with increasing forecast lead time. This concurrent weakening indicates that lower wind speeds are associated with reduced radial geopotential gradients, demonstrating that the large-scale DDM patterns of geopotential and winds are consistent with the expectations from gradient-wind balance. Nevertheless, Fig.~\ref{fig:figure7} also clearly highlight a progressive underestimation of Poly’s storm intensity in the Bris predictions at longer lead times, manifested by weaker winds and a less sharply defined geopotential minimum. It is evident that, although the NWP forecast skill also decreases with lead time, it maintains closer agreement with the radial profiles observed in the analysis.



\subsection{Continuity equation}

\begin{figure}[h!]
    \centering
    \includegraphics[width=0.9\textwidth]{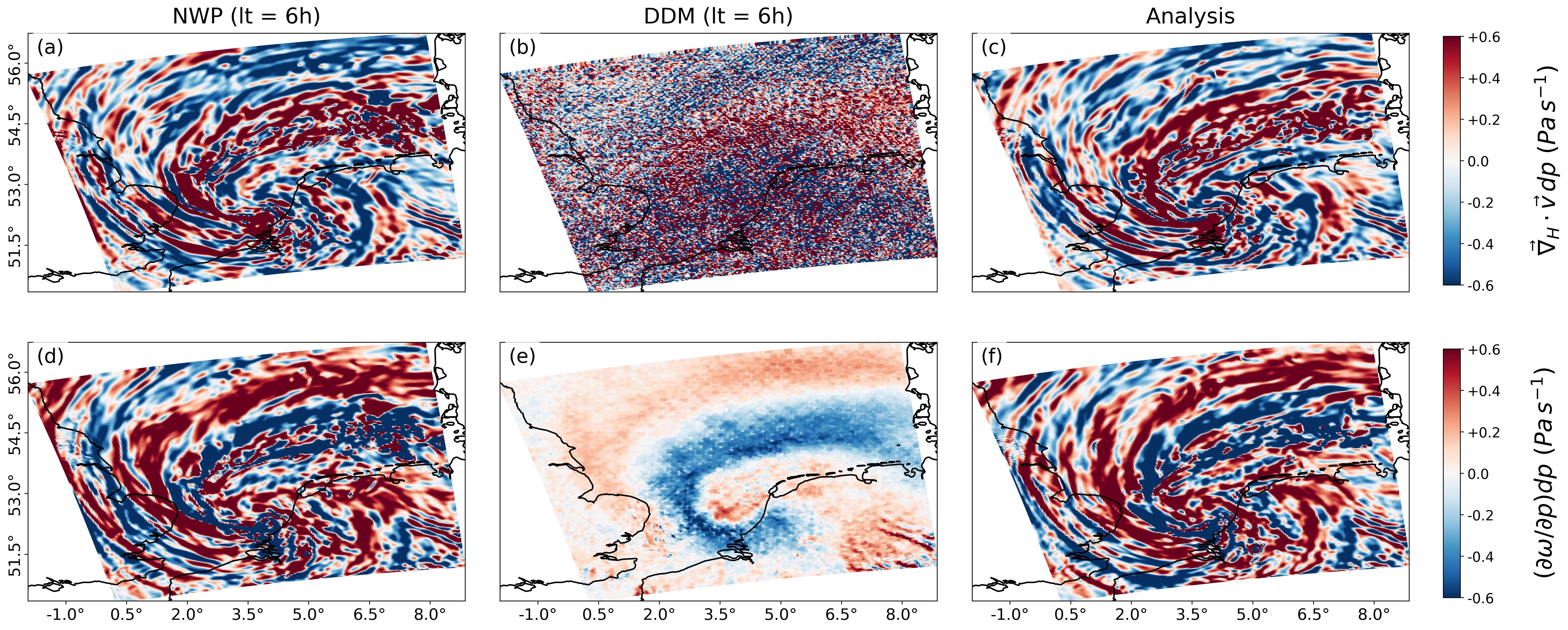}
    \caption{\small (a–c) vertically integrated horizontal wind divergence over the 500–400 hPa layer; (d–f) vertical difference of the vertical velocity  over the same 500–400 hPa layer. The maps show a 250 × 250 grid-point subdomain centered (roughly) on the storm’s LPC at 06 UTC on 5 July 2023. Panels refer to NWP (left) and DDM (centre) +6 h forecasts, and the corresponding analysis field (right).}
    \label{fig:figure8}
\end{figure}

Fig.~\ref{fig:figure8} shows that, in the two physics-based fields (NWP and analysis), a clear degree of anticorrelation is present: regions of negative vertical gradients of vertical velocity (blue patches) systematically coincide with areas of positive horizontal wind divergence (red patches), and viceversa. This behaviour is consistent with the continuity equation and reflects a physically coherent dynamical structure. In contrast, the DDM predictions fail to reproduce this anticorrelation, as the horizontal divergence field does not exhibit a coherent spatial relationship with the vertical velocity gradients. This qualitative assessment is further supported by a quantitative analysis based on the Pearson correlation coefficient; the result can be found in the MSc thesis by \citet{Pasquini2025} connected to the present work. 

At larger spatial scales, the Bris model still exhibits a moderate negative correlation between the two variables. A broad, spiral-shaped blue structure associated with the warm conveyor belt of Poly is clearly visible in the vertical gradient of vertical velocity (Fig.~\ref{fig:figure8}e). Consistently, in the same region, a pronounced cluster of scattered red dots is partially discernible in the horizontal wind divergence field (Fig.~\ref{fig:figure8}b). This suggests that the large-scale dynamics of the windstorm are, at least to some extent, adequately represented. However, at smaller spatial scales, this anticorrelation progressively weakens and eventually disappears. The horizontal wind divergence predicted by Bris becomes highly irregular and erratic, exhibiting systematic noise that disrupts any coherent dynamical structure.

\subsection{Fine-scale noise and spectral analysis}

\begin{figure}[h!]
    \centering
    \includegraphics[width=1.0\textwidth]{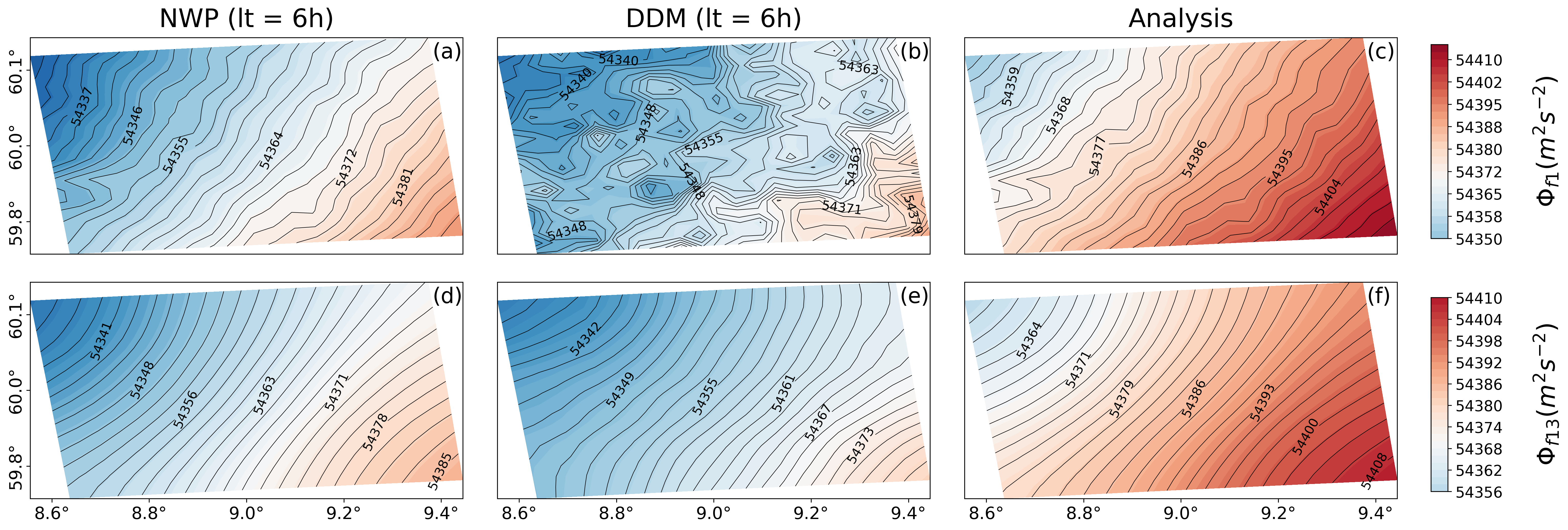}
    \caption{\small Inset ($\sim$50 $\times$ 50 km) of the 500hPa geopotential field for (a-c) raw (unsmoothed) (a) NWP, (b) DDM +6h predictions, and (c) analysis fields at 06 UTC on 5 July 2023, and for (d-f) the corresponding fields after application of the f13 smoothing filter. The selected area corresponds to a small square located in the northwestern portion of the 443 $\times$ 443 subdomain where conditions at the considered time are expected to be close to geostrophic balance.}
    \label{fig:figure9}
\end{figure}

The analysis of the physical balances reveals clear signatures of erratic and unphysical noise in Bris predictions, amplified at higher resolutions (see, e.g., the isogeopotential lines in Fig.~\ref{fig:figure6}b and the divergence field in Fig.~\ref{fig:figure8}b). A closer inspection of the 500hPa geopotential field further exposes these spurious small-scale features in the DDM (Fig.~\ref{fig:figure9}b), which contrast to the much smoother fields in the NWP model (Fig.~\ref{fig:figure9}a) and analysis (Fig.~\ref{fig:figure9}c). This noise affects \textit{all} output variables of Bris (including wind components, temperature and specific humidity) and is particularly easy to discern in the geopotential field, due to its intrinsically smooth structure in NWP forecasts and analyses. Importantly, after applying the f13 smoothing filter to the geopotential field, the spatial patterns in the DDM better resemble those of the NWP and analysis outputs and the noise disappears (Fig.~\ref{fig:figure9}d-f).

\begin{figure}[h!]
    \centering
    \includegraphics[width=0.45\textwidth]{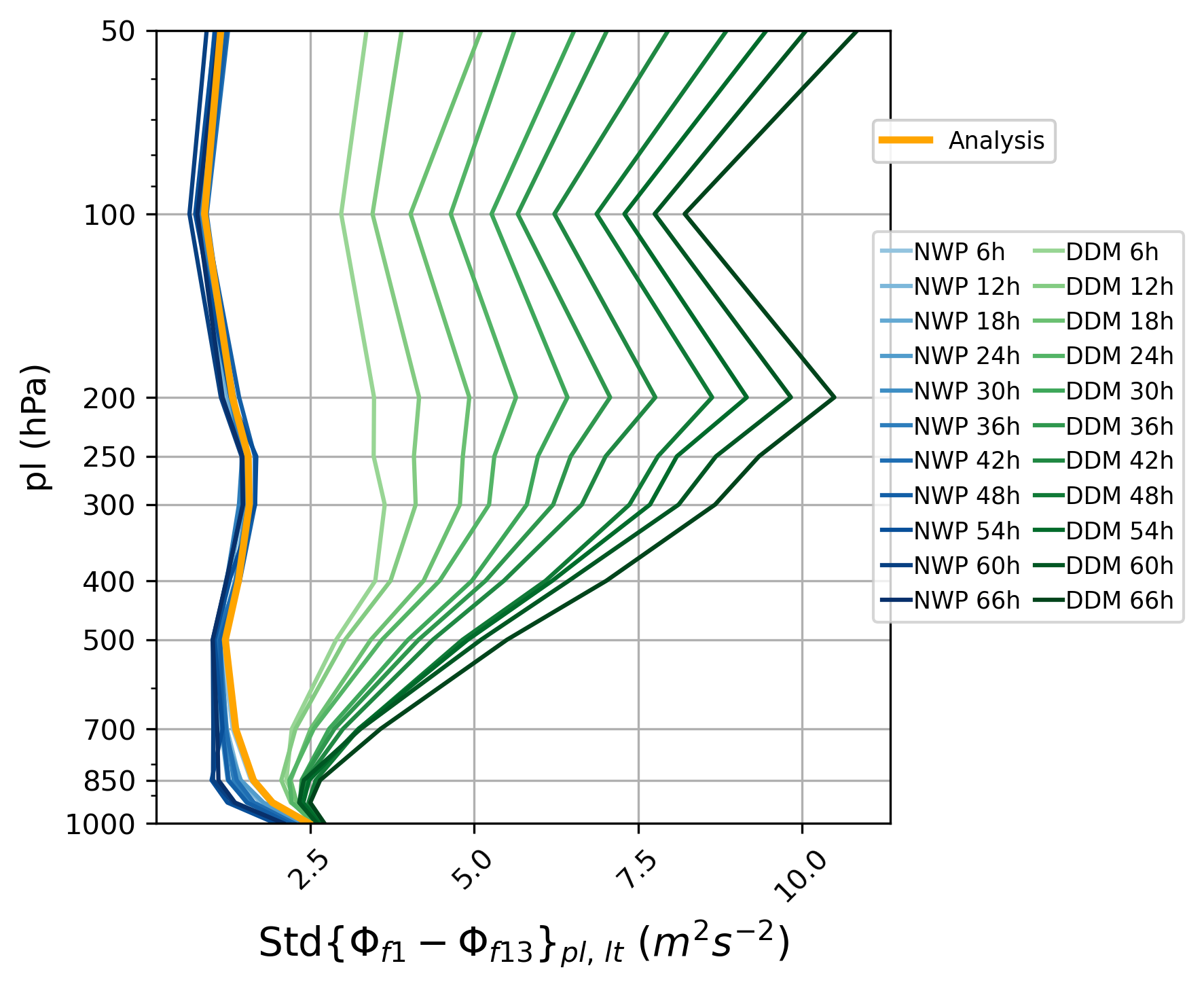}
    \caption{\small Vertical profiles of the standard deviation of the differences between the unsmoothed ($\Phi_{f1}$) and smoothed ($\Phi_{f13}$) geopotential fields. Differences $\Phi_{f1} - \Phi_{f13}$ are computed over the all 443 $\times$ 443 grid-point subdomain considered in this study, for multiple lead times, at 06 UTC on 5 July 2023. Results are shown for the NWP model (blue), the DDM (green), and the analysis (orange).}
    \label{fig:figure10}
\end{figure}

\begin{figure}[h!]
    \centering
    \includegraphics[width=1.0\textwidth]{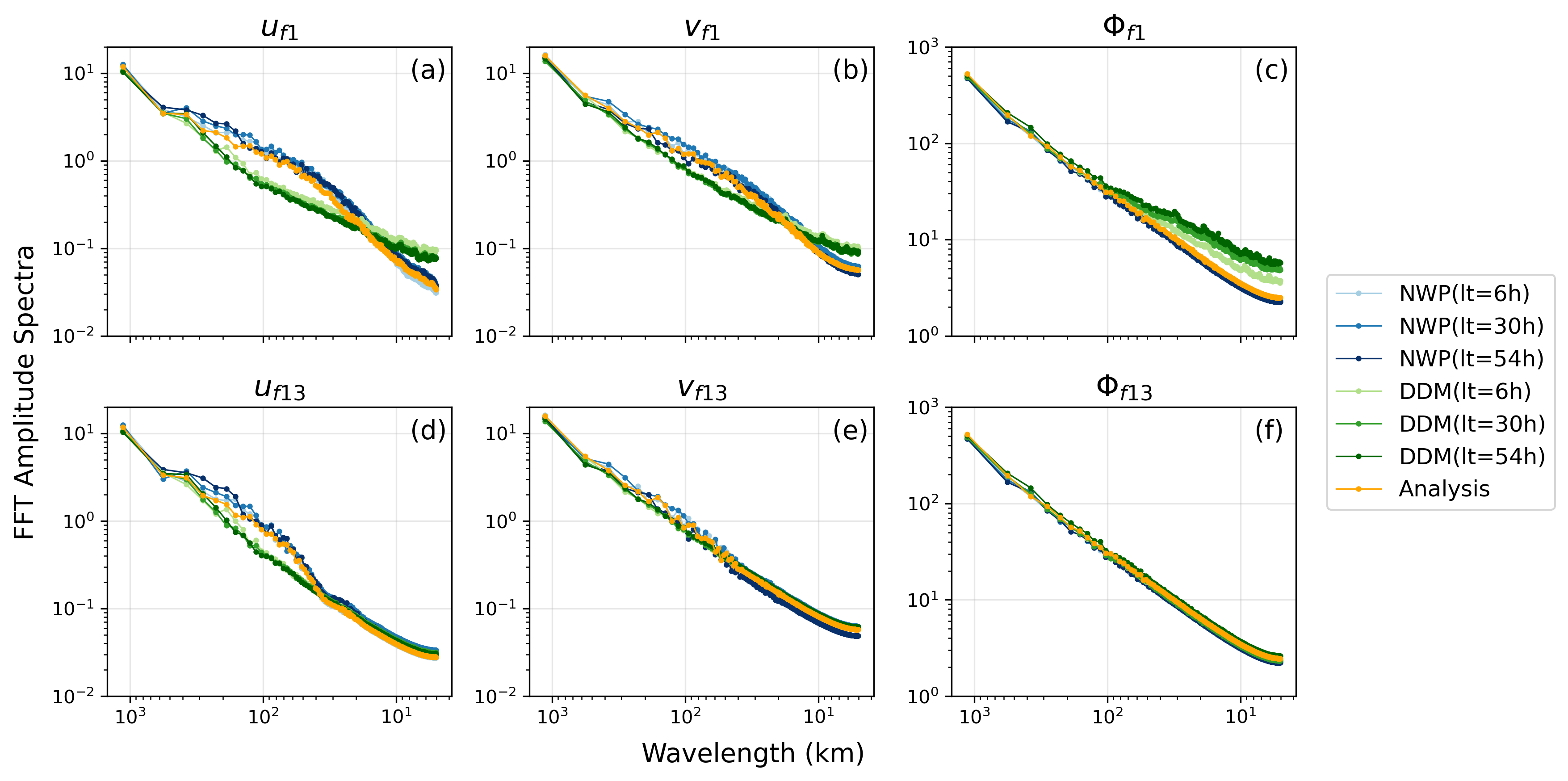}
    \caption{\small Fast Fourier Transform (FFT) amplitude spectra as a function of wavelength (a–c) for the unsmoothed (f1) (a) zonal wind, (b) meridional wind, and (c) geopotential at 500hPa, and (d-f) for the corresponding smoothed (f13) fields. Colored lines denote the different datasets and lead times listed in the legend. The spectra are shown for 18 UTC on 5 July 2023 and refer to the 443 $\times$ 443 geographical subdomain used in this study; the spatial mean of each field was subtracted before computing the spectra.}
    \label{fig:figure11}
\end{figure}

In the absence of fine-scale noise (as in the NWP and analysis data), the distribution of differences between the unsmoothed ($\Phi_{f1}$) and smoothed ($\Phi_{f13}$) geopotential fields is expected to be narrowly centered, since the 13$\times$13 neighborhood mean closely approximates the unsmoothed value. In contrast, fine-scale noise produces a broader $\Phi_{f1} - \Phi_{f13}$ distribution, reflecting larger deviations from the surrounding average. The standard deviation (Std) of this difference can therefore be used as a proxy for the magnitude of fine-scale noise. As shown in Fig.~\ref{fig:figure10}, the DDM exhibits a substantially larger Std at higher altitudes (lower pressures) and longer lead times, while smaller values occur near the surface and at shorter lead times. This behaviour is not present in the NWP and analysis data, suggesting that these unrealistic small-scale variations are intrinsic to the Bris output and intensify both with forecast lead time and altitude.

Additional insights are provided by the spectral analysis in Fig.~\ref{fig:figure11}. Overall, the analysis spectra show better agreement with the MEPS outputs than the Bris predictions. We see that, at large wavelengths (30-300 km), the DDM spectra of the horizontal wind components (u and v) are markedly underestimated. In contrast, the amplitude spectra of the horizontal winds and the geopotential are significantly amplified at smaller wavelengths (<20 and <100 km, respectively). The underestimation of wind amplitudes at large scales is consistent with the well-known smoothing and blurring effects associated with the use of an MSE-based loss function. Conversely, the overrepresentation of short-wavelength variability is likely linked to the erratic, small-scale variations recently identified in the stretched-grid DDM forecasts.

Fig.~\ref{fig:figure11}a and \ref{fig:figure11}b illustrate that these seemingly competing features—large-scale smoothing and small-scale noise—can coexist within the same predictions. Fig.~\ref{fig:figure11}d-f show that applying the f13 smoothing filter effectively suppresses the unrealistic variability in the fields. The $\Phi_{f13}$ spectra in Fig.~\ref{fig:figure11}c further indicate that the noise amplitude increases with forecast lead time, consistent with the trends shown in Fig.~\ref{fig:figure10}. Repeating the spectral analysis for the corresponding surface variables yields similar results, suggesting that the erroneous small-scale variations persist near the surface, albeit at generally lower intensity, likely because of the higher weight of the surface variables in the loss function.

\section{Discussion}
\label{sec:discussion}

In this article, we show that, despite the rarity of the case-study event, the stretched-grid data-driven model retains the RMSE skill previously reported by \citet{Nipen2024}. Furthermore, Bris demonstrates a clear ability to forecast the occurrence of the Poly storm, which is an encouraging outcome given the event’s exceptional impact, its red-alert classification by KNMI \citep{Kees2024}, and the well-documented difficulties of ML-based weather models in representing extreme events \citep{Olivetti2024, Pasche2024}. The introduction of this stretched-grid model therefore points to the potential emergence of a real alternative to traditional deterministic high-resolution NWP systems. 

At the same time, the examination highlights several clear limitations of the model when its performance is compared to MEPS. We observe clear signs of large-scale smoothing in predicted meteorological fields, a trait that seems typical of DDMs employing MSE as the loss function \citep[e.g.,][]{Bi2022,Lam2023,Lang2024}. In the context of windstorm Poly, a related negative effect of this smoothing is that, especially at longer lead times, the strong wind speeds near the LPC of the cyclone are largely absent in the Bris predictions. From an operational perspective, this represents a notable shortcoming, as such intense winds are often critical to anticipate the potential impact of a storm. Additionally, the DDM longer-term predictions generally exhibit weaker pressure gradients and a significantly shallower LPC, resulting in an overall underestimation of the cyclone’s intensity.

This research reveals two additional critical limitations of Bris that warrant careful consideration. First, the model struggles to produce predictions that respect fundamental physical balances (e.g., see unrealistic deviations from geostrophic balance in Fig.~\ref{fig:figure4},~\ref{fig:figure5}, and see findings for the continuity equation presented in Fig.~\ref{fig:figure8}). Second, all DDM output variables are affected by pronounced fine-scale noise at high resolution (e.g., Fig.~\ref{fig:figure9}), which results in exaggerated energies in the spectra at the finest scales (Fig.~\ref{fig:figure11}). Both the unrealistic deviations and the fine-scale noise were observed to increase systematically with longer lead times and at higher altitudes (e.g., Figs.~\ref{fig:figure4} and \ref{fig:figure10}), suggesting that these two issues are closely interrelated.

Taking the geostrophic balance as an illustrative example, this connection becomes evident. Fine-scale noise introduces unrealistic spatial gradients in the geopotential field (e.g., erroneous Jacobians: $\frac{\partial \Phi}{\partial x}$ and $\frac{\partial \Phi}{\partial y}$), which directly affect the computation of ageostrophic wind components (Eq.~\ref{eq:geostrophic_wind},~\ref{eq:geostrophic_imbalance}). Locally, these wrong gradients may even align with the Coriolis force, substantially disrupting geostrophic equilibrium and resulting in physically implausible ageostrophic winds. More broadly, this indicates that any atmospheric balance governed by horizontal spatial gradients is likely to be disrupted when such erratic small-scale variability affects the predicted fields.

Another important finding of this study is that applying a smoothing filter to the DDM predictions substantially improves the consistency with fundamental atmospheric balances (e.g., Fig.~\ref{fig:figure4},~\ref{fig:figure5}) and yields spatial patterns that are more consistent with those from numerical forecasts and analysis fields (Fig.~\ref{fig:figure9}). This improvement likely arises because smoothing suppresses the fine-scale noise present in the raw DDM output, which is the primary source of disruption of physical balances. For example, when examining the gradient wind balance (Fig.~\ref{fig:figure6}), we did not observe the same highly unrealistic imbalances evident in the geostrophic balance. Although the gradient wind balance also depends on horizontal geopotential gradients, the geopotential field was first fitted to the original data prior to computing the gradient winds. This fitting procedure inherently introduces smoothing, effectively suppressing fine-scale noise and resulting in a more physically consistent balance.

Specifically, we found that an f13 filter—corresponding to averaging each grid point over a 13 × 13 surrounding window (32.5 $\times$ 32.5 km)—is sufficient to effectively suppress the background noise (e.g., Fig.~\ref{fig:figure9},~\ref{fig:figure11}). Fields smoothed with this filter exhibit substantially improved agreement with the examined physical balances. This is a particularly relevant result, considering that Bris was trained globally over a 43-year period at a comparable spatial resolution (i.e., the same resolution as ERA5). Additional evidence of improved physical consistency at larger scales is provided by the analysis of the gradient wind balance in the cyclone-relative reference frame (Fig.~\ref{fig:figure6},~\ref{fig:figure7}) and by the continuity equation (Fig.~\ref{fig:figure8}). These diagnostics show that some key aspects of the expected large-scale dynamics of cyclone Poly are realistically represented in the DDM predictions. 

However, while the DDM succeeds at reproducing the balanced state of the NWP and the analysis fields, it is much less successful at reproducing the imbalances, which we interpret as being mostly physical in the NWP model (Fig.~\ref{fig:figure3},~\ref{fig:figure6}). We therefore cannot use the DDM output to study physical processes over time, as the transient responses are fundamentally tied to these imbalances.

\section{Conclusions}

We evaluated the ability of the stretched-grid data-driven model to produce physically consistent and realistic forecasts during the severe extratropical cyclone Poly. Our analysis revealed shortcomings that extend beyond this specific case, highlighting a general limitation of the stretched-grid DDM in reproducing the characteristic spatial patterns of meteorological fields present in its training data at high resolution. The regional forecasts generated by Bris are affected by systematic fine-scale noise, which causes an excess of energy at small scales in the spectra, and leads to the breakdown of fundamental atmospheric balances.

Both the magnitude of this noise and the resulting physical inconsistencies increase with lead time of forecasts, consistent with the expected degradation of skill. The noise and balances disruption also become more pronounced with altitude. This vertical dependence may stem from the operationally motivated decision to formulate a loss function which gives greater weight to near-surface levels while giving relatively less weight to the upper atmosphere. 

A clear explanation underlying such unrealistic high variability at finer-scales remains unclear. To the best of the authors' knowledge, physical inconsistencies of comparable magnitude have not been reported in lower-resolution DDMs with similar architecture and loss function, suggesting that the Bris stretched-grid approach may indeed play a role. The issue, for instance, may arise from the imbalance in the training process between coarse- and fine-resolution (Bris is pre-trained on 43 years of global ERA5 data at 31 km resolution and is subsequently fine-tuned using only 3.3 years of 2.5 km resolution operational analyses \citep{Nipen2024}). The encoder--processor--decoder GNN architecture itself may also contribute to the observed unrealistic fine-scale variability. In the processor mesh of the GNN, spatial relationships are based on graph nodes and edges whose structure does not preserve the regular Cartesian grid spacing typically assumed when computing horizontal gradients.

Importantly, a carefully designed modification of the loss function can also help reduce these artifacts. We refer to the probabilistic version of Bris \citep{Nordhagen2025}, where a loss was formulated to account not only for local point-wise errors in physical space but also for discrepancies in the representation of spatial scales in the spectral domain and in the distribution of energy across different frequencies. This approach was shown to improve the fine-scale structure of the predicted fields, and as a result of the work here presented, we emphasize that such improvements are crucial for ensuring that DDM forecasts better respect atmospheric balances.  

Finally, we also note that improvements are needed to enhance Bris’s reliability in representing the intensity of extreme windstorms such as Poly. These limitations underscore the broader challenges faced by deterministic high-resolution DDMs in operational forecasting. Future work should also investigate the performance of probabilistic stretched-grid DDMs under extreme weather, to provide a more comprehensive assessment of their potential and limitations for real-world applications.

\section*{\MakeUppercase{Author contributions}}
F.P. designed and conducted the study and wrote the first draft of the manuscript. 
M.B. supervised the work and provided key conceptual insights that were essential to the success of the study. 
B.F. and N.T. substantially supported the study, contributing significantly to both the study setup and the writing of the manuscript. 
M.S. originally conceived the analysis and initiated the project, supervising the work from its conceptual design to the writing of the manuscript. 

\section*{\MakeUppercase{Acknowledgements}}
We sincerely thank Thomas Nipen (MET Norway) for providing the Bris model output dataset, which was essential to this study. We are also grateful to Șerban Vădineanu (KNMI), Sophie Buurman (KNMI), Harun Kivril (KNMI) and Wim de Rooij (KNMI) for valuable discussions that contributed to improving the quality of this work.

\section*{\MakeUppercase{CONFLICT OF INTEREST STATEMENT}}
The authors declare no conflicts of interest.

\section*{\MakeUppercase{Data availability and code}}
MEPS forecasts and analyses are publicly available \citep{MEPSarchive}, and the code used in this study can be accessed at \url{https://github.com/FrancescoPasquini25/Supplementary_codes_for_Bris_evaluation}.



\renewcommand{\refname}{\MakeUppercase{References}}

\newpage
\section*{\MakeUppercase{Supplementary Figures}}

\renewcommand{\thefigure}{S\arabic{figure}}
\setcounter{figure}{0}

\begin{figure}[h!]
    \centering
    \includegraphics[width=0.6\textwidth]{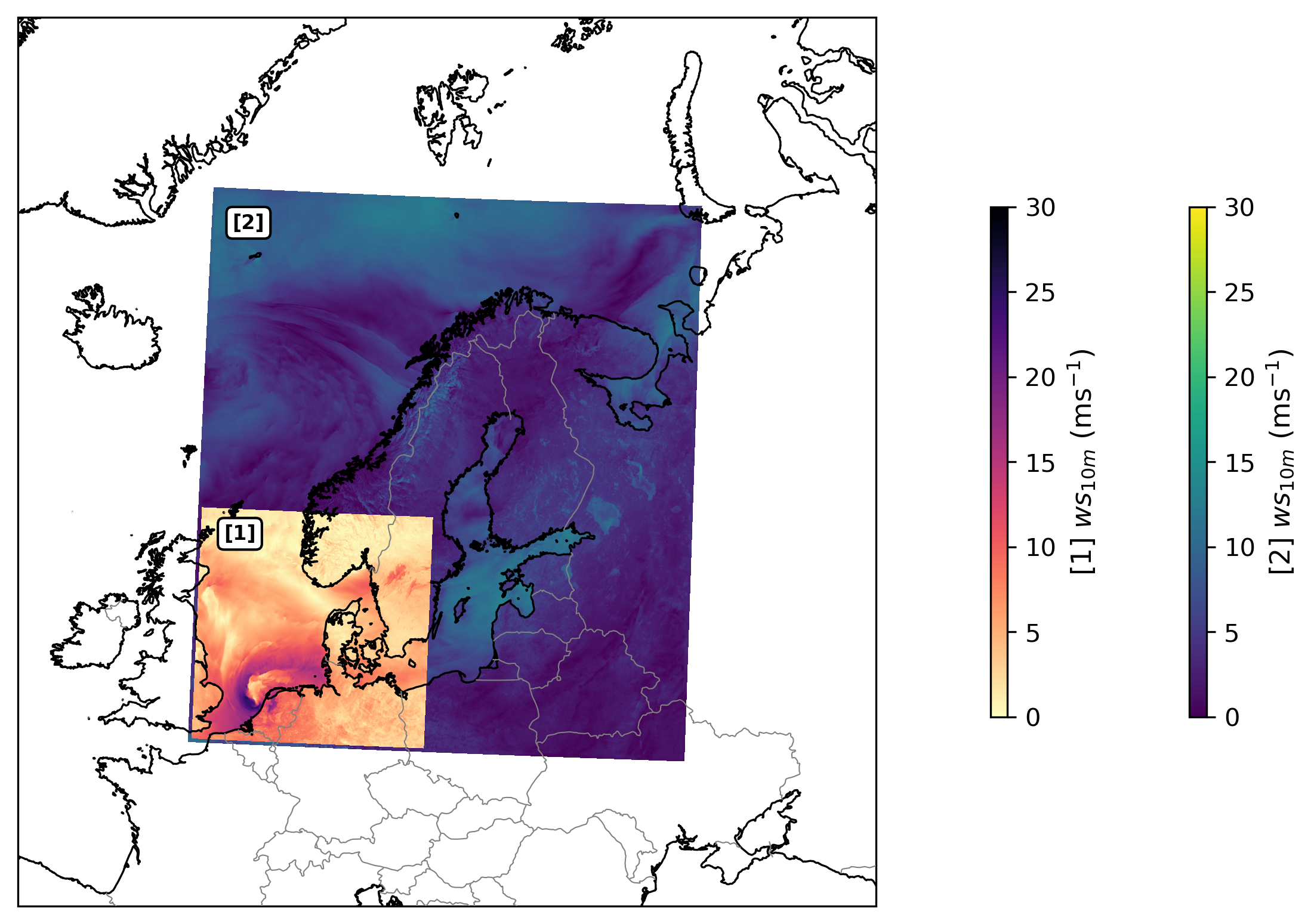}
    \caption{\small 10m wind speed maps at 06 UTC on 5 July 2023 in [1] the 443 × 443 subdomain used in this study and in [2] the full regional high-resolution domain of the Bris and MEPS models. In both [1] and [2], the horizontal grid spacing is 2.5 km; in Bris the global grid spacing is 0.25° (approximately 31 km) outside these regions.}
\end{figure}

\begin{figure}[h!]
    \centering
    \includegraphics[width=0.95\textwidth]{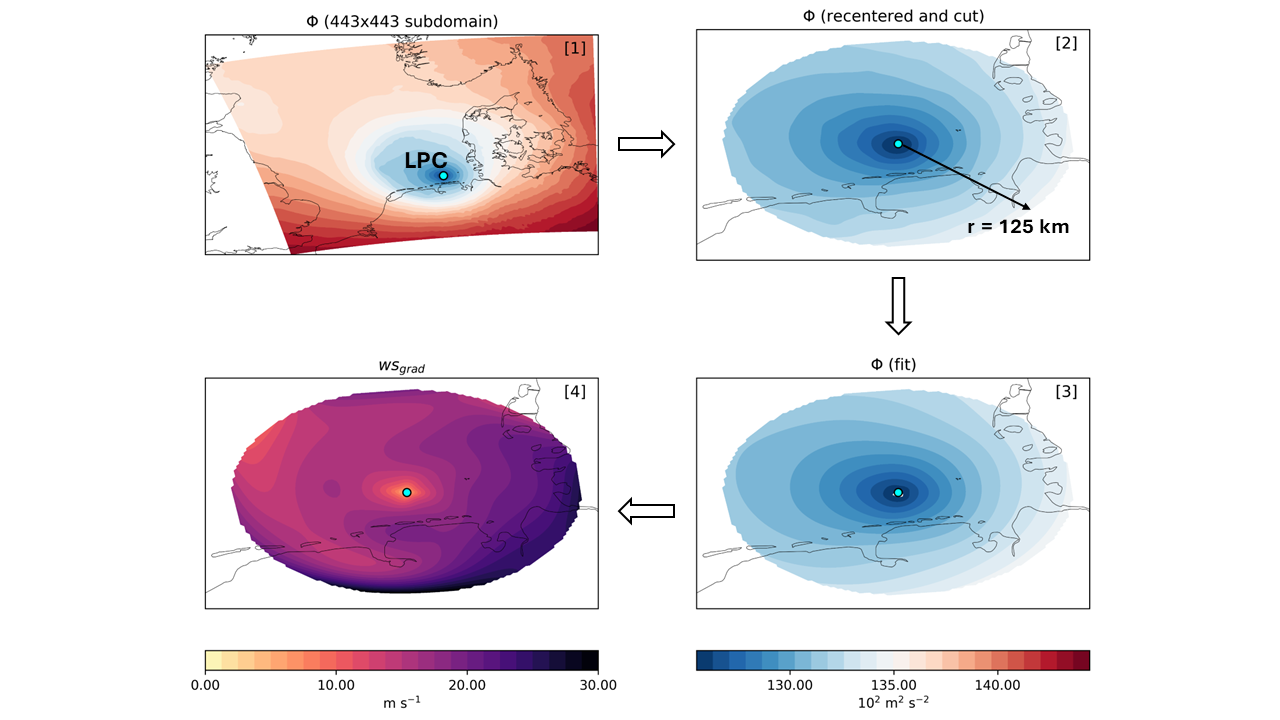}
    \caption{\small Step-by-step procedure for deriving the gradient wind in Storm Poly within a cyclone-centered reference framework. [1] Geopotential at 850hPa at 12UTC on 5 July 2023 over the 443$\times$443 subdomain; the green dot marks the minimum geopotential location (LPC). [2] Recentered geopotential field around the predicted LPC, with values retained only within a 125~km radius (others set to NaN). [3] Geopotential fit using Eq.~\eqref{Phi_fit}. [4] Gradient wind computed from Eq.~\eqref{eq:gradient_wind_balance}.}
\end{figure}


\end{document}